\documentclass[10pt,twocolumn]{article}

\usepackage{titlesec}

\titleformat*{\section}{\large\bfseries}
\titleformat*{\subsection}{\bfseries}

\usepackage[english]{babel}
\usepackage[utf8]{inputenc}
\usepackage{CJKutf8}

\usepackage[normalem]{ulem}

\usepackage{amsfonts,amsmath,amssymb,graphicx}
\usepackage{cuted}  
\usepackage[usenames,dvipsnames]{xcolor}
\usepackage{setspace}
\usepackage{hyperref}
\usepackage{amsmath}

\usepackage{pdfpages}

\newcommand{\be}{\begin{eqnarray}}
\newcommand{\ee}{\end{eqnarray}}

\newcommand{\E}{\mathcal{E}}

\usepackage{color,balance,xspace,verbatim,ifthen,engord}

\usepackage{amsmath,wasysym,amsthm,marvosym,stackengine}

\usepackage{booktabs,colortbl,diagbox,multirow,tabularx,tablefootnote}

\usepackage{caption}
\usepackage{subcaption}
\usepackage{graphicx,epsfig,epstopdf,wrapfig}
\graphicspath{{./figcam/}} 
\DeclareGraphicsExtensions{.pdf,.mps,.png,.jpg,.eps,.PNG,.JPG}
\usepackage{algorithm,algorithmic}

\usepackage{url}

\usepackage{tikz}

\newcommand*\circled[1]{\tikz[baseline=(char.base)]{\node[shape=circle,draw,inner sep=0.5pt] (char) {#1};}}

\def\eg{\textit{e.g.,}\hspace{1mm}}
\def\ie{\textit{i.e.,}\hspace{1mm}}

\usepackage{courier}

\usepackage{enumitem}

\usepackage{xcolor,soul}
\usepackage{float}
\usepackage{enumitem}

\usepackage{pf2}

\newcommand{\remove}[1]{}

\definecolor{lightgray}{gray}{0.6}
\definecolor{lightblue}{rgb}{0.9,0.9,1}
\definecolor{aqua}{rgb}{0.0, 1.0, 1.0}

\newcommand{\region}{region\xspace}
\newcommand{\regions}{regions\xspace}

\newcommand{\cluster}{dual cluster\xspace}
\newcommand{\clusters}{dual clusters\xspace}

\newcommand{\csize}{size\xspace}

\newcommand{\diagram}{Fowler diagram\xspace}

\newcommand{\papertitle}{An interpretation of Union-Find Decoder on Weighted Graphs}
\newcommand{\personnel}{Yue Wu, Namitha Liyanage, and Lin Zhong\\ Department of Computer Science, Yale University, New Haven, CT}

\begin{document}

\title{\papertitle}

\author{\personnel}

\maketitle

\begin{abstract}

Union-Find (UF) and Minimum-Weight Perfect Matching (MWPM) are popular decoder designs for surface codes.
The former has significantly lower time complexity than the latter but is considered somewhat inferior, in terms of decoding accuracy.
In this work we present an interpretation of UF decoders that explains why UF and MWPM decoders perform closely in some cases: the UF decoder is an approximate implementation of the blossom algorithm used for MWPM.
This interpretation allows a generalization of UF decoders for weighted decoding graphs and explains why UF decoders achieve high accuracy for certain surface codes.

\end{abstract}

\section{Introduction}
\label{sec:intro}

Fast and accurate quantum error correction is necessary for fault-tolerant quantum computing. Surface codes have emerged as one of the leading choices for quantum error correction. 
A surface code interleaves data and ancilla qubits on a surface such that an error in a data quit will impact the measurement outcome of its neighboring ancillas. The decoder's job is to determine the error pattern, i.e., which data qubits experience errors, based on the syndrome, i.e., measurement outcomes of all ancillas. 

Union-find (UF)~\cite{delfosse2020linear} and Minimum-Weight Perfect Matching (MWPM) are two popular decoder designs. Both leverage graph representations of surface code and its syndromes. 
A UF decoder works on the model graph where vertices correspond to measurement qubits and edges correspond to data qubits with weights determined by the error probability.
An MWPM decoder works on the syndrome graph, a fully-connected graph generated from the model graph in which vertices correspond to nontrivial measurement outcomes and an edge corresponds to minimum-weight paths in the model graph. As its name suggests, an MWPM decoder finds a minimum-weight perfect matching of the syndrome graph and uses it as its guess for the error pattern. Union-Find is known to be much faster than MWPM decoders but in general, achieves lower decoding accuracy.

The primary contribution of this work is to reveal a hidden link between UF and MWPM decoders.
By adapting UF decoders to work on the syndrome graph, we show that its working principles approximate those of the blossom algorithm, a highly optimized algorithm that finds an MWPM.
Both UF and Blossom algorithm decompose a syndrome graph into non-overlapping subgraphs. 
Both start with the syndrome graph with each subgraph including a single vertex and grow these subgraphs such that each can be ``solved'' on its own.
While the blossom algorithm finds an MWPM for each subgraph, a UF decoder finds a logical equivalent to a PM for it.
We show that under many circumstances, a UF decoder and the blossom algorithm may end up decomposing the syndrome graph in the same way and their solutions for subgraphs may be logically equivalent.

Once revealed, the link between UF and MWPM decoders not only allows us to improve the accuracy of UF decoders and generalize them for weighted decoding graphs but also explain why UF decoders achieve similar accuracy as MWPM decoders for certain surface codes, such as the XZZX surface code~\cite{bonilla2021xzzx} when noise is infinitely biased and measurements are perfect.
Specifically, we devise a more general UF decoder that works for surface codes without assuming identical error probability for data qubits.

We presented the interpretation and its implications at~\cite{yue2022aps}. Since then we have learned that others had reached similar interpretations independently~\cite{google-interpretation}.  At the same time, we have also learned that many are still unaware of the link between the UF and blossom algorithms. Therefore, our motivations in writing this article are two. First, to share the interpretation with the community in hope to foster the development of fast quantum error decoders. Second, to formalize the interpretation as rigorous as we could.  

In the rest of the paper, we provide necessary background in \S\ref{sec:background}; we describe how a UF decoder approximates the blossom algorithm in \S\ref{sec:interpretation}.
Using this interpretation, we explain why a UF decoder behaves the same as an MWPM decoder for two examples in \S\ref{sec:example}. 
We derive weighted union-find decoders based on the interpretation in \S\ref{sec:weighted_uf}. We have open-sourced their implementations at~\cite{qec-playground}.
Mathematical details of proofs can be found in the Appendix.

\section{Background}
\label{sec:background}

\subsection{Surface code}
\label{sec:background_surface}

\begin{figure*}[ht]
    \renewcommand*\thesubfigure{(\alph{subfigure})}  
    \centering
    \begin{subfigure}{.3\linewidth}
        \centering
        \includegraphics[width=0.7\linewidth]{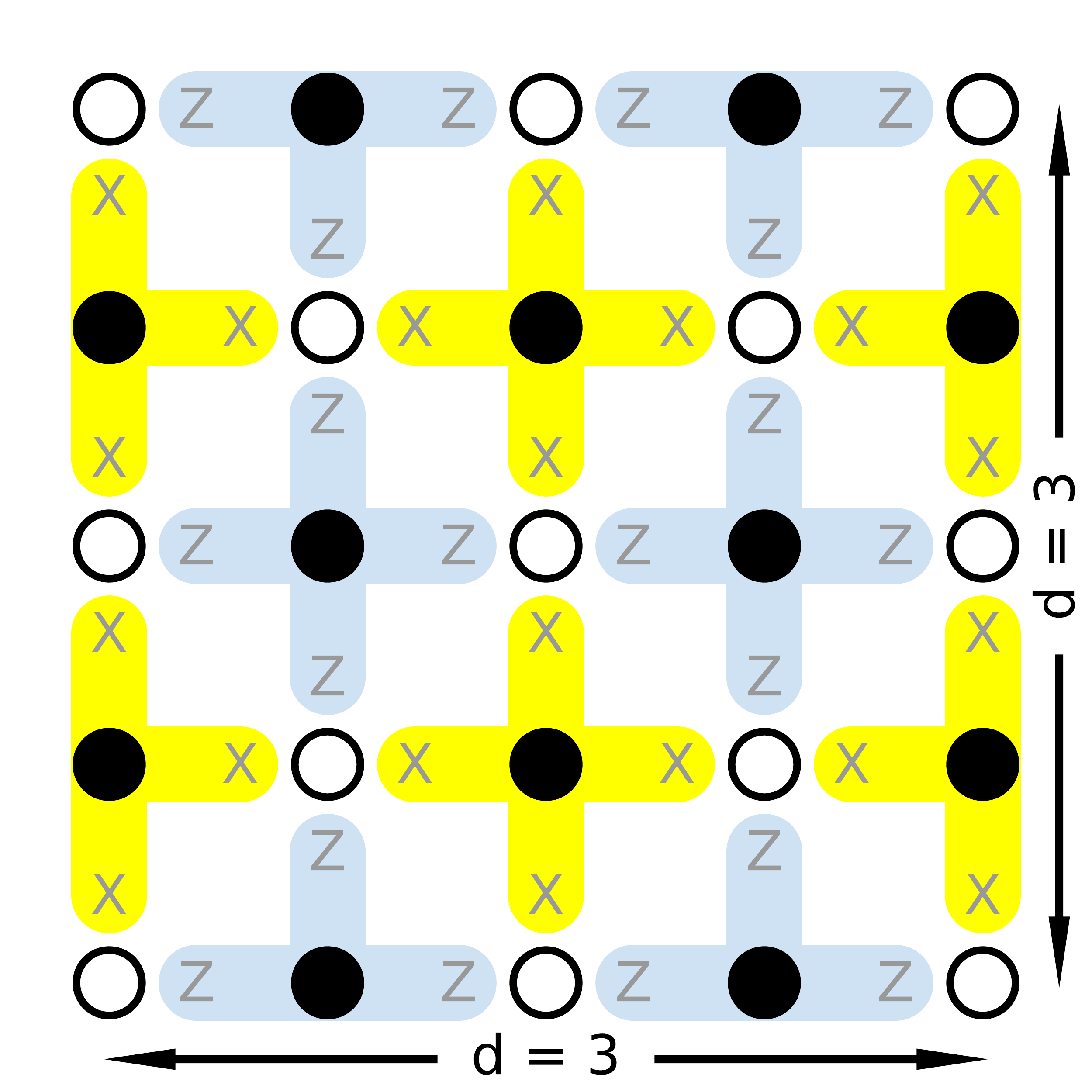}
        \caption{}
        \label{fig:basic_CSS}
    \end{subfigure}
    \begin{subfigure}{.3\linewidth}
        \centering
        \includegraphics[width=0.7\linewidth]{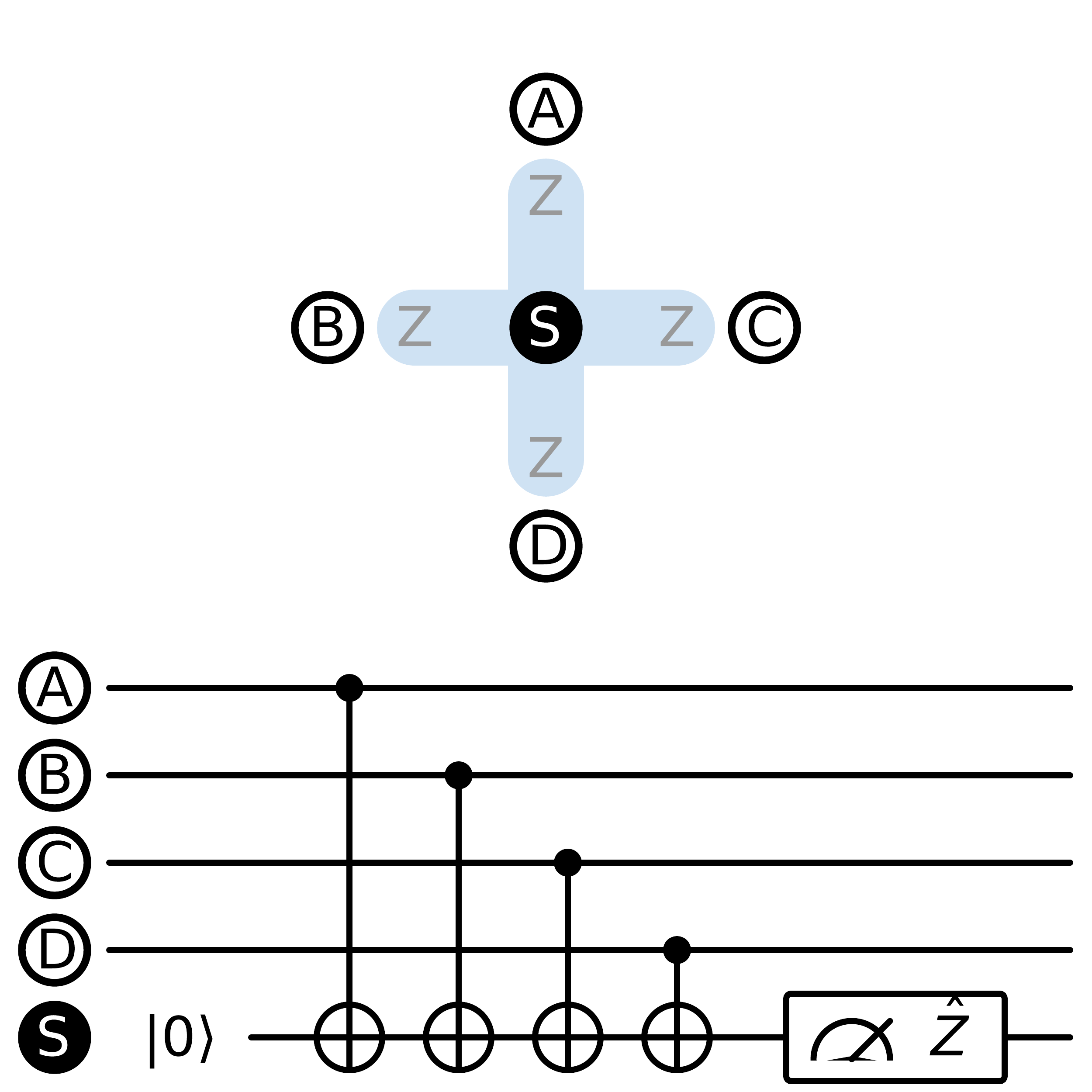}
        \caption{}
        \label{fig:CSS_Z_circuit}
    \end{subfigure}
    \begin{subfigure}{.3\linewidth}
        \centering
        \includegraphics[width=0.7\linewidth]{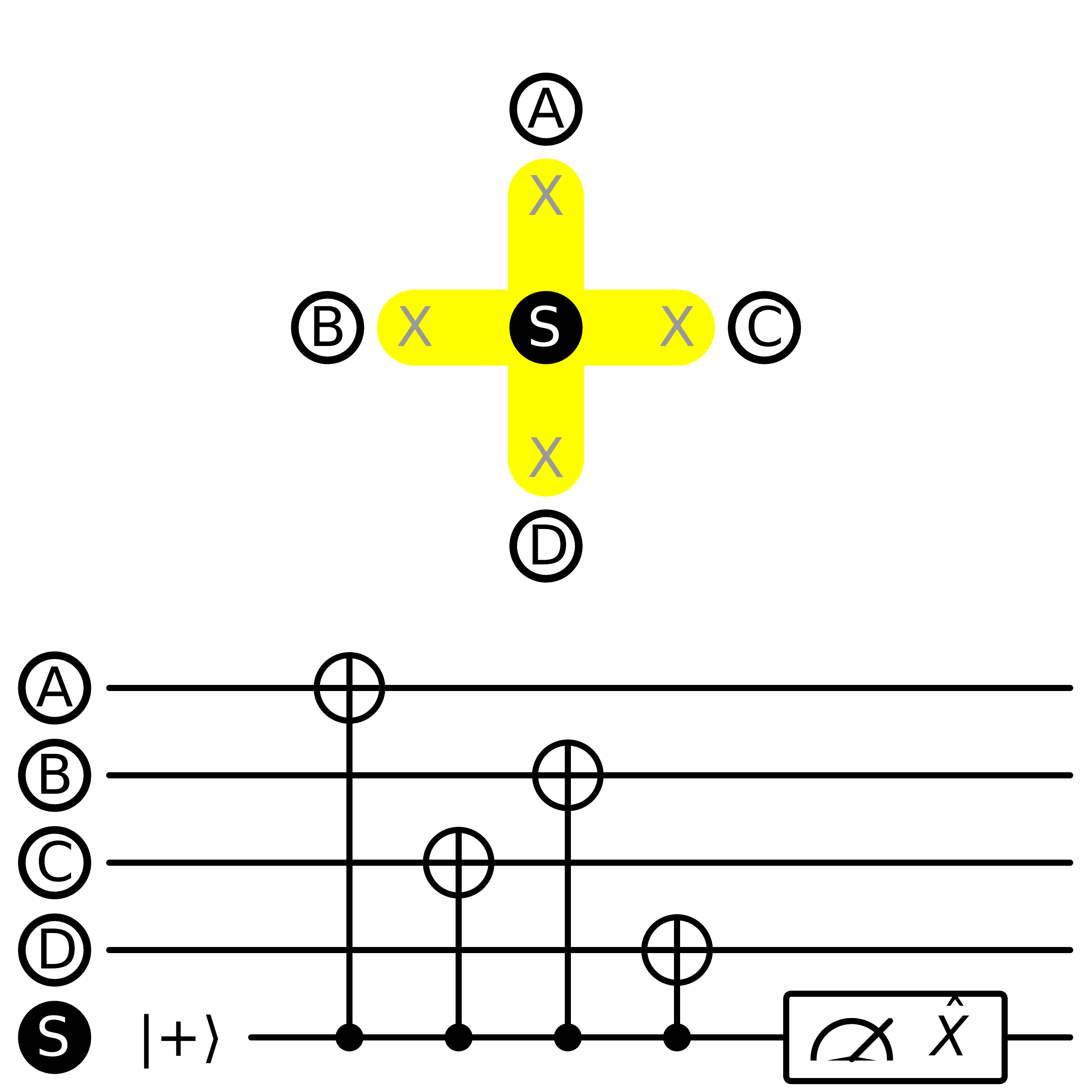}
        \caption{}
        \label{fig:CSS_X_circuit}
    \end{subfigure}
    \caption{(a) : distance-3 CSS surface code. The data qubits are shown in white circles and the Z-type and X-type ancillas are shown in black circles. (b) and (c) : Measurement circuit of Z-type and X-type ancillas. Excluding the ancillas in the border, each Z-type and X-type ancilla interacts with 4 adjacent data qubits. Z-type and X-type ancillas have different gate orders to prevent certain type of hook error~\cite{yoder2017surface, tomita2014low}.}
    \label{fig:surfacecode}
\end{figure*}

A surface code is a lattice of interleaved ancilla and data qubits arranged on a surface in which each ancilla qubit entangles with neighboring data qubits.
In this work, we focus on surface codes with open boundaries. That is, the surface is a bounded plane.
Measuring an ancilla qubit is equivalent of performing a \emph{stabilizer} operation on its entangled data qubits. 
We assume the stabilizer measurements are noiseless for simplicity, and will generalize to noisy stabilizer measurements in \S\ref{ssec:generalize3d}.
\autoref{fig:surfacecode} shows an instance of the popular CSS surface code.
The stabilizer is either $\otimes^4 X$ (X-type ancilla) or $\otimes^4 Z$ (Z-type ancilla) on adjacent data qubits as shown in \autoref{fig:CSS_Z_circuit} and \autoref{fig:CSS_X_circuit}, respectively.
An X (or Z) error in a data qubit will flip the measurement outcomes between +1 and \textminus 1 of its Z-type (or X-type) ancillas.

A qubit, logical or physical, can be represented by a vector in a 2-dimensional Hilbert space. 
The array of $N$ data qubits is therefore represented by a vector in a $2^{N}$-dimensional Hilbert space.
An error-correction code identifies a subspace of this huge space, called \emph{code space}.
Each state in the code space corresponds to a state in the 2-dimensional Hilbert space, i.e., a \emph{logical} state. That is, a code space state encodes a \emph{logical} qubit. 
In an error-free world, as the array functions as a logical qubit, it will remain in the code space. 
An error in a data qubit will bring the array outside the code space. 

Because direct measurement is destructive, QEC relies on the measurement outcomes of ancilla qubits, i.e., the \emph{syndrome}, to detect errors and then correct them, to bring the array back the code space, hopefully to the same \emph{logical} state. If the array ends up in a different logical state, a logical error happens. 

Both an error pattern, denoted by $\mathcal{E}$, and a correction, $C$, can be considered as an operator $\hat{\mathcal{E}}$ and $\hat{\mathcal{C}}$ that changes the states of the corresponding data qubits.
Any error pattern or multi-qubit operator can be written as a linear combination of Pauli errors (operators).
Hence, we only look at decoding a tensor product of identity operators $\hat{I}$ or Pauli operators ($\hat{X}$, $\hat{Y}$, and $\hat{Z}$), which directly leads to $\hat{\mathcal{E}} \hat{\mathcal{E}} = \hat{I}$.
That is, $C=\mathcal{E}$ corrects the error pattern $\mathcal{E}$.
However, there are usually more than one correction $C$ that could avoid the logical state change under $\mathcal{E}$. We denote the set of such corrections as $\mathbf{C}_\mathcal{E}$.

\begin{figure*}[!ht]
    \renewcommand*\thesubfigure{(\alph{subfigure})}  
    \centering
    \begin{subfigure}{.16\linewidth}
        \centering
        \includegraphics[width=1\linewidth]{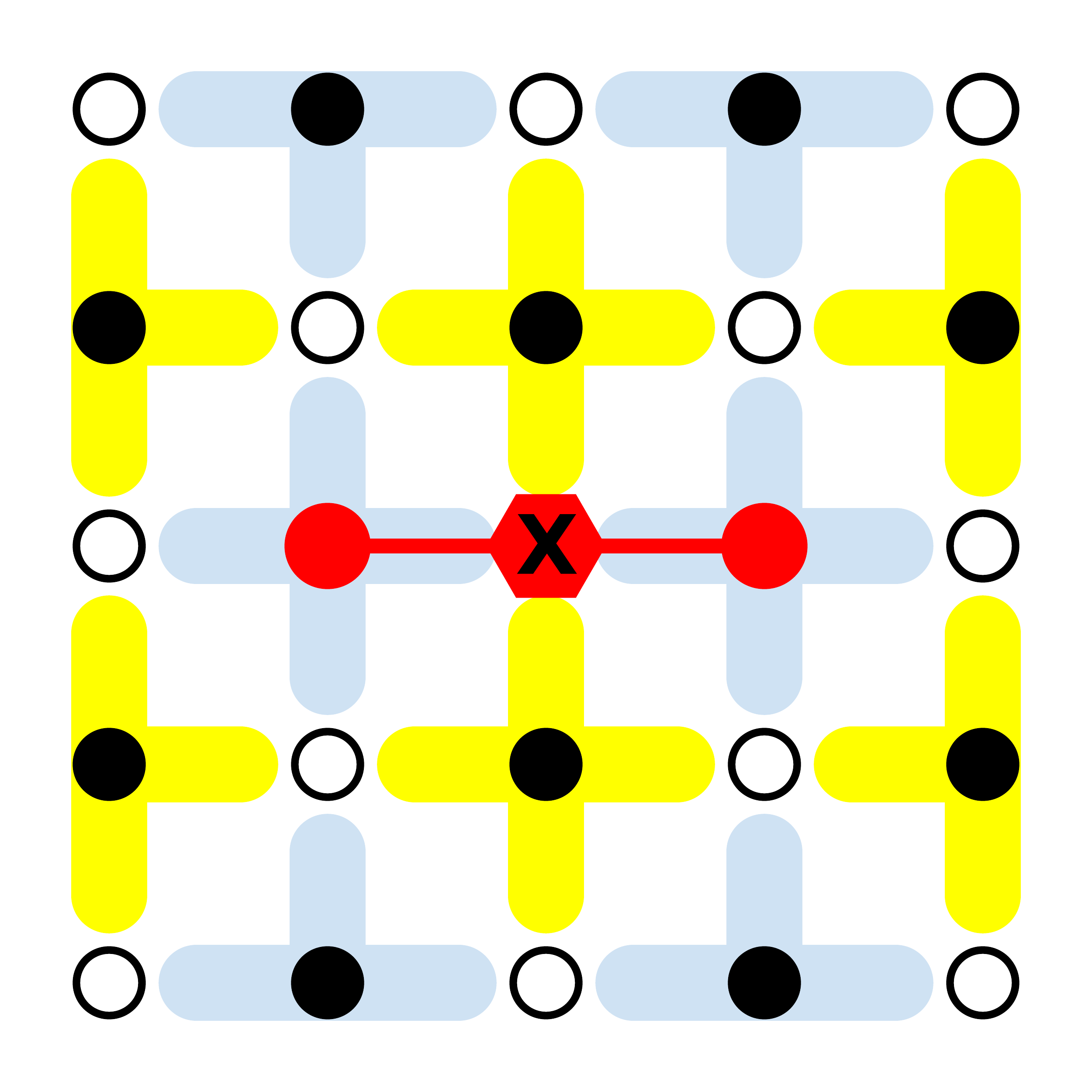}
        \caption{}
        \label{fig:single_X_error}
    \end{subfigure}
    \begin{subfigure}{.16\linewidth}
        \centering
        \includegraphics[width=1\linewidth]{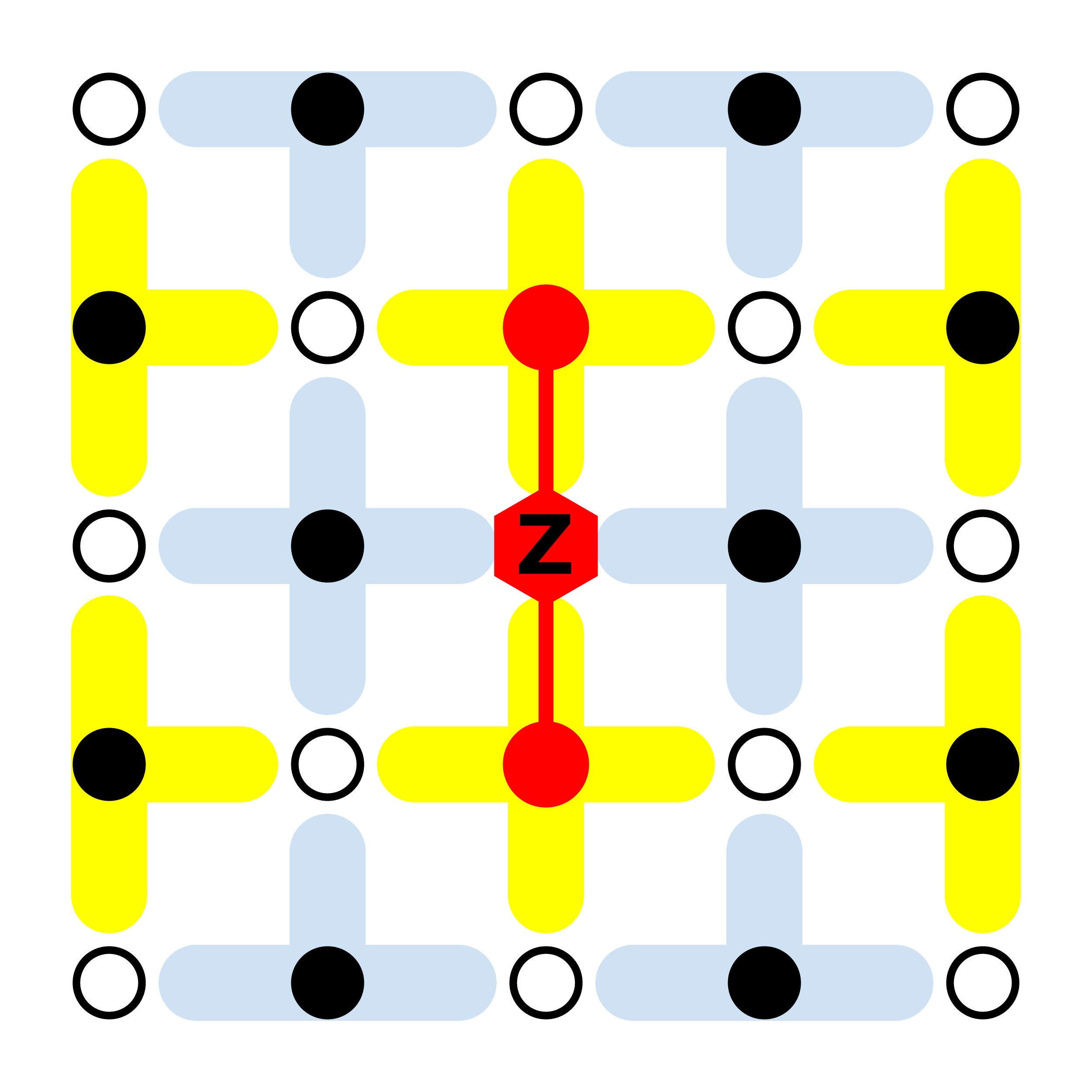}
        \caption{}
        \label{fig:single_Z_error}
    \end{subfigure}
    \begin{subfigure}{.16\linewidth}
        \centering
        \includegraphics[width=1\linewidth]{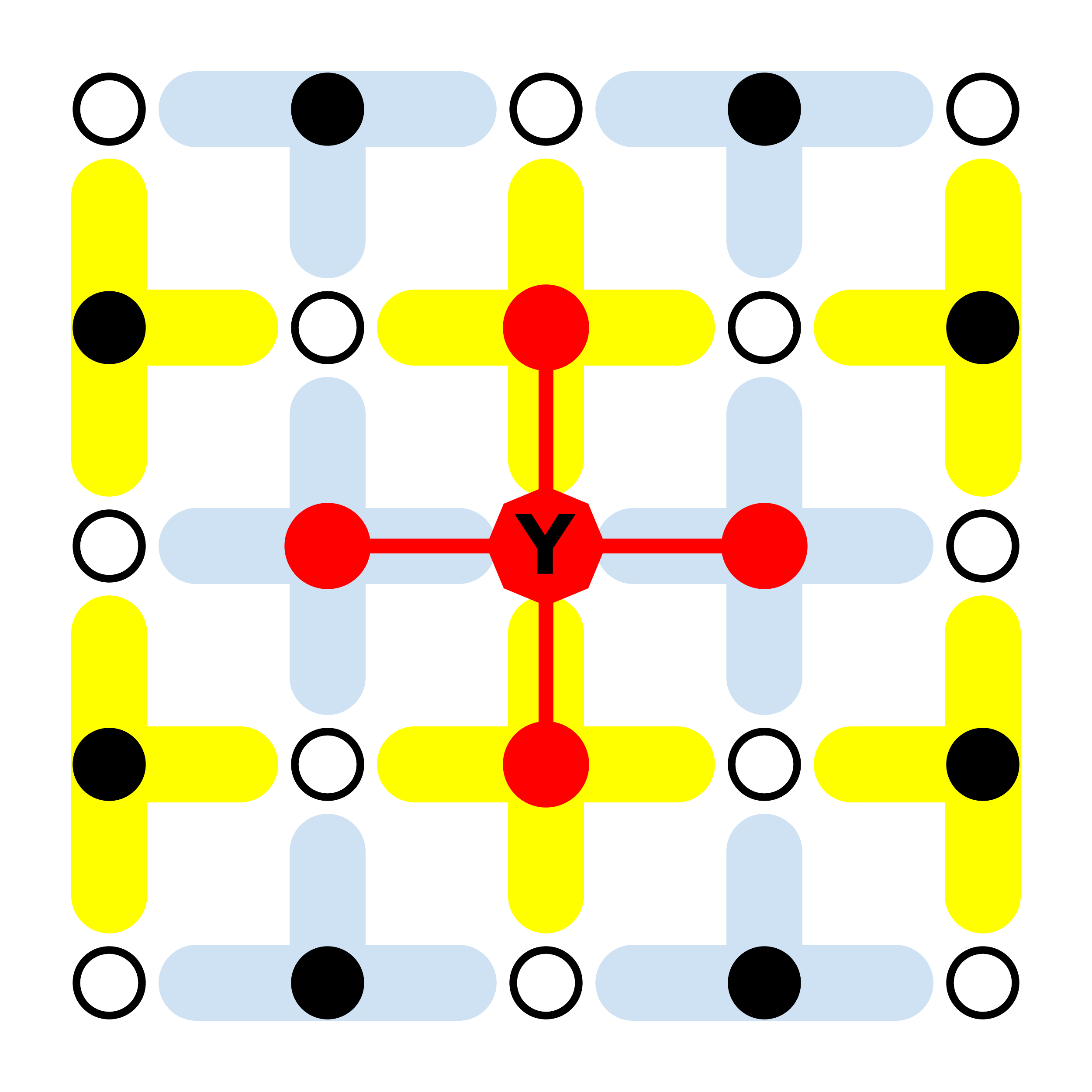}
        \caption{}
        \label{fig:single_Y_error}
    \end{subfigure}
    \begin{subfigure}{.16\linewidth}
        \centering
        \includegraphics[width=1\linewidth]{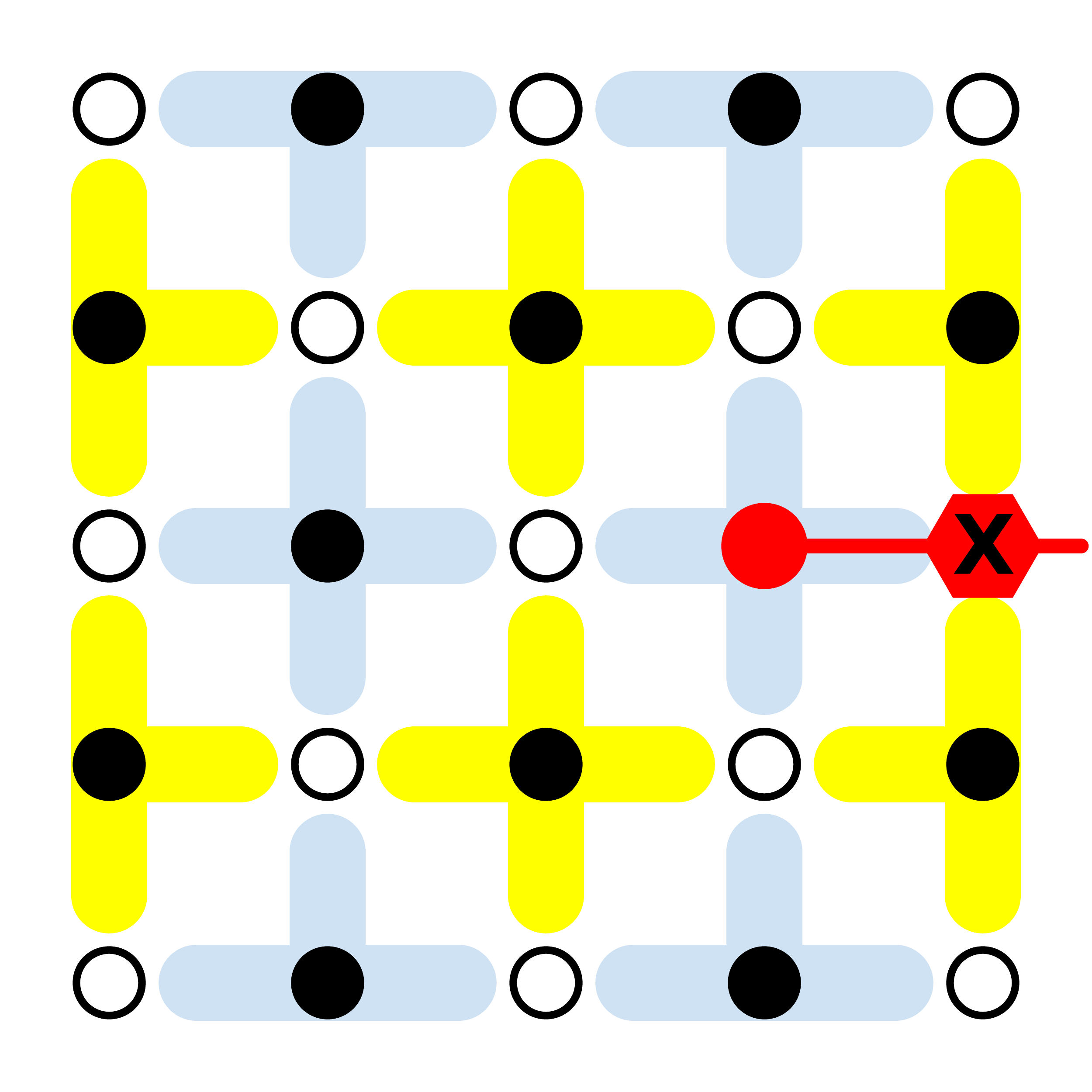}
        \caption{}
        \label{fig:single_X_boundary_error}
    \end{subfigure}
    \begin{subfigure}{.16\linewidth}
        \centering
        \includegraphics[width=1\linewidth]{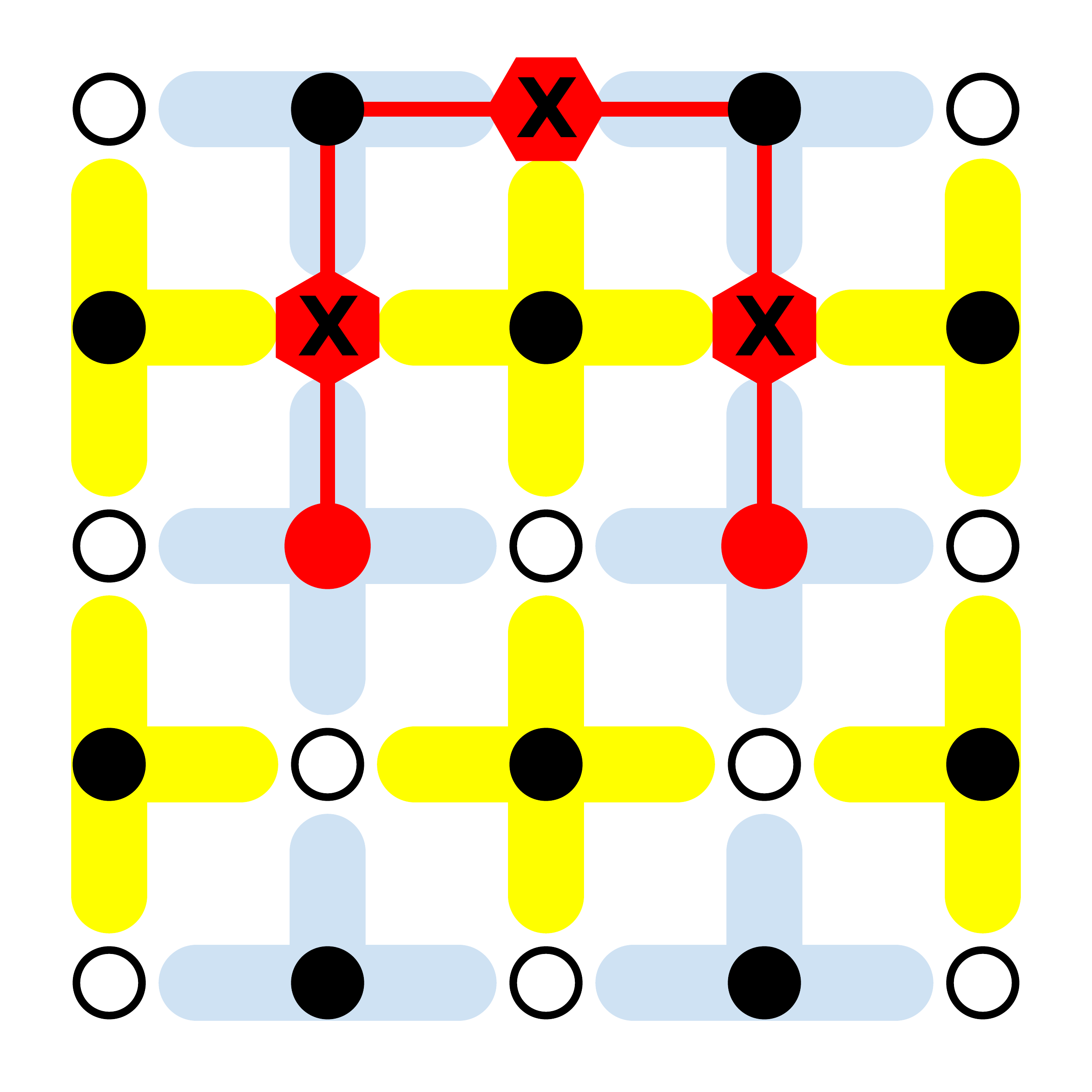}
        \caption{}
        \label{fig:X_chained_error}
    \end{subfigure}
    \begin{subfigure}{.16\linewidth}
        \centering
        \includegraphics[width=1\linewidth]{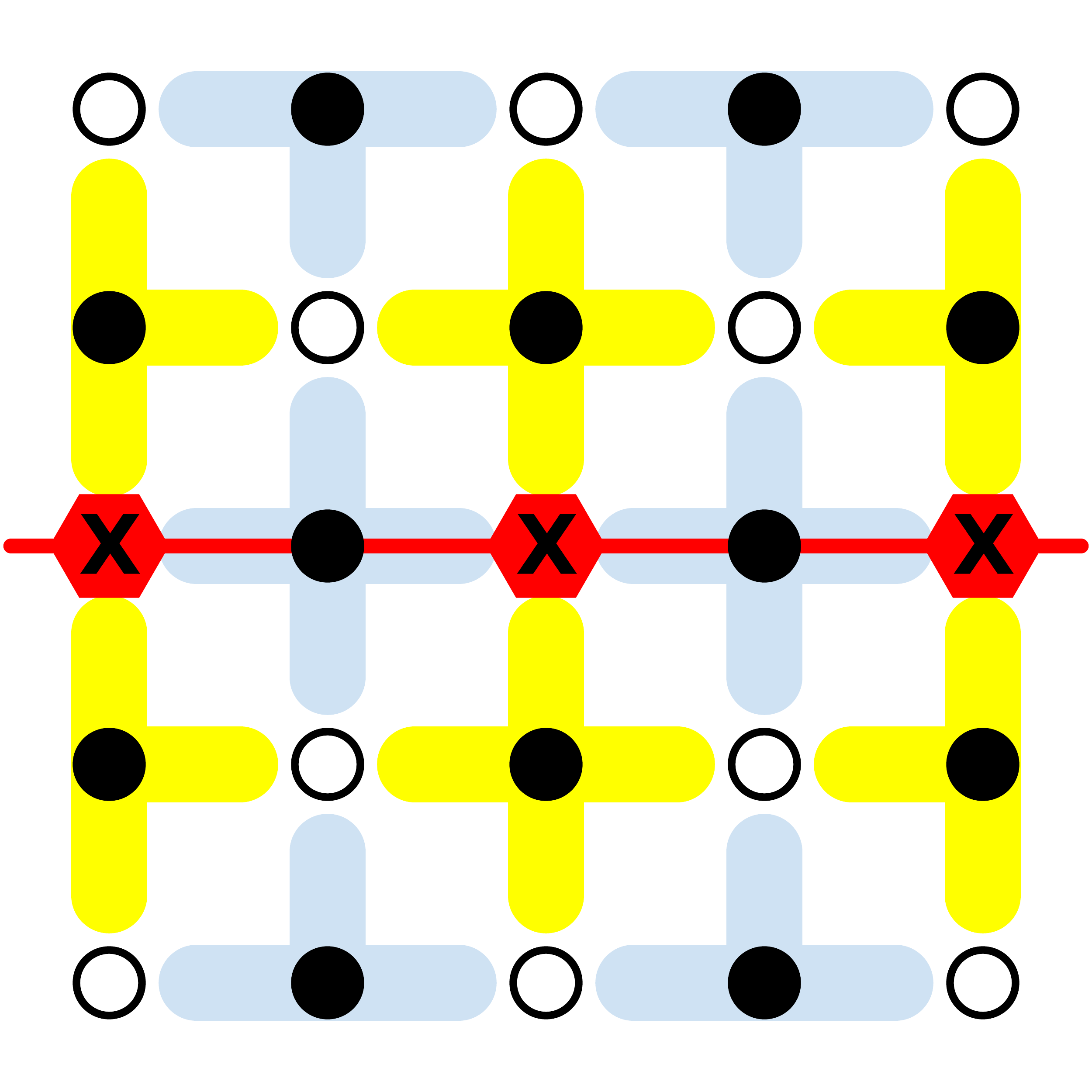}
        \caption{}
        \label{fig:X_logical_error}
    \end{subfigure}
    \caption{Visualization of multiple error patterns on d = 3 surface code. The physical qubit errors are marked by X and Z. Ancillas reporting nontrivial measurements are shown in red. The red lines are to visualize error chains. (a) isolated X error (b) isolated Z error (c) isolated Y error (d) X error generating single nontrivial measurement (e) error chain of three X errors (f) error chain introducing a logical error. Note that even though (a) and (e) are different error patterns, they produce the same syndrome.}
    \label{fig:error-patterns}
\end{figure*}

Error correction is challenging because different error patterns may have the same syndrome. Two classes of error patterns have no syndrome at all, which means they are not detectable and can be considered as the \emph{logical operators} of the surface code.
In the first class, errors of the same type (X or Z) form a closed circle. 
Such a pattern (and its corresponding operator) does not affect the logical state of the surface code.
That is, it is a trivial logical operator, i.e., identity.
Apparently, any $C\in \mathbf{C}_{\mathcal{E}}$, $\hat{C} \hat{\mathcal{E}}$ must be a trivial logical operator. 
In the second class, errors of the same type (X or Z) form a continuous chain connecting the respective boundaries, as shown \autoref{fig:X_logical_error}.
Such a pattern (and its corresponding operator) changes the logical state of the surface code.
That is, it is a nontrivial logical operator (X or Z) of the surface code.
We note these classes exist for general surface codes but involve more nuanced topology of the surface and error patterns~\cite{bombin2013introduction}.

\subsection{Decoding and its challenges} 
Generally speaking, a decoder starts with the syndrome, denoted by $S$, and the error model, denoted by the probability distribution of error patterns $P(\mathcal{E})$, and computes a correction, denoted by $C$, which is an operator on the surface code.
The error model $P(\mathcal{E})$ is determined by the code and the hardware; it can be acquired offline. In contrast, the syndrome $S$ is measured at runtime.

Given the syndrome $S$, we can compute the probability of not having a logical error if correction $C$ is applied as
\begin{equation}
P(C,S)= \sum\limits_{\mathcal{E}| C\in{\mathbf{C}_\mathcal{E}}} P(\mathcal{E}|S)
\label{eq:probability}
\end{equation}
Both $P(\mathcal{E}|S)$ and $\mathbf{C}_\mathcal{E}$ can be computed offline.

Then we can compute the best correction $C(S)$, in terms of lowest probability of logical error as

\begin{equation}
    \begin{aligned}
C(S)=\arg \max\limits_C P(C,S)\\=\arg \max\limits_C \sum\limits_{\mathcal{E}| C\in{\mathbf{C}_\mathcal{E}}} P(\mathcal{E}|S)
\label{eq:coset}
\end{aligned}
\end{equation}

Note that the optimal correction $C(S)$ is generally not unique, and we denote the set of the best corrections for $S$ with $\mathbf{C}(S)$.

\textbf{Challenges}~~The above process, however, faces scaling challenges in both storage and computation. 
(\textit{i}) First, there are about $4^{2d^2}$ different error patterns given about $2d^2$ data qubits each can have no error (identity $I$) or a Pauli $X$, $Z$ or $Y$ error.
As a result, a complete error model $P(\mathcal{E})$ requires $4^{2d^2}$ entries.
(\textit{ii}) Second, given $S$ and $P(\mathcal{E})$, the computational complexity for $\mathbf{C}(S)$ is $O(d^2 (4^{2d^2})^2)$ using \autoref{eq:coset}.

\textbf{Lookup table decoder} tackles Challenge (\textit{ii}) above by pre-computing an optimal $C(S)$ for every possible syndrome $S$ offline, storing them in a table, and looking it up at runtime, with constant decoding time.
It does not address Challenge (\textit{i}), which makes the offline computation non-scalable. 
Moreover, the table has one entry per syndrome and there are about $2^{2d^2}$ possible syndromes.
Even for $d=5$, the table will have over a quadrillion ($10^{15}$) entries.

\section{Graph Based Decoding}
\label{sec:graph}
In contrast, graph-based decoding tackles both challenges, in two stages.
The first stage constructs a graph that approximates the \emph{error model} $P(\mathcal{E})$.
The second stage then uses this graph and the syndrome to compute a correction. Both stages can employ heuristics to simplify the computation, finding a ``good'' correction within polynomial time.
Because the error model is usually known given the hardware and code, i.e., offline, and the syndrome is only observed at runtime, i.e., online, only the second stage needs to be computed at runtime.

\subsection{Graph Construction}
\label{sec:graph_construction}
The first stage constructs a weighted graph that represents an approximation of the error model $P(\mathcal{E})$, which we call the \textit{model graph}.
We first describe how to construct the model graph assuming that errors happen independently and only with data qubits. We will discuss how the graph can be extended to consider other types of errors in \S\ref{ssec:generalize3d}.
Under the independence assumption, one can reduce $P(\mathcal{E})$ into  
\begin{equation}
\begin{split}
    P(\mathcal{E}) &= \prod_{\mathcal{E}_i \in \mathcal{E}} P(\mathcal{E}_i) \prod_{\mathcal{E}_i \notin \mathcal{E}} \left(1-P(\mathcal{E}_i)\right) \\
                   &= \prod_{\mathcal{E}_i \in \mathcal{E}} \frac{P(\mathcal{E}_i)}{1 - P(\mathcal{E}_i)} \prod_{\mathcal{E}_i} \left(1-P(\mathcal{E}_i)\right) \\
                   &\propto \prod_{\mathcal{E}_i \in \mathcal{E}} \frac{P(\mathcal{E}_i)}{1 - P(\mathcal{E}_i)}
    \label{eq:independence}
\end{split}
\end{equation}
where $\mathcal{E}_i$ indicates the $i$th independent error.
If the independence assumption is true for single data qubit Pauli X and Z errors, one can construct two separate graphs to represent the X and Z error models. 

\begin{figure*}[ht]
    \renewcommand*\thesubfigure{(\alph{subfigure})}  
    \centering
    \begin{subfigure}{.24\linewidth}
        \centering
        \includegraphics[width=0.76\linewidth]{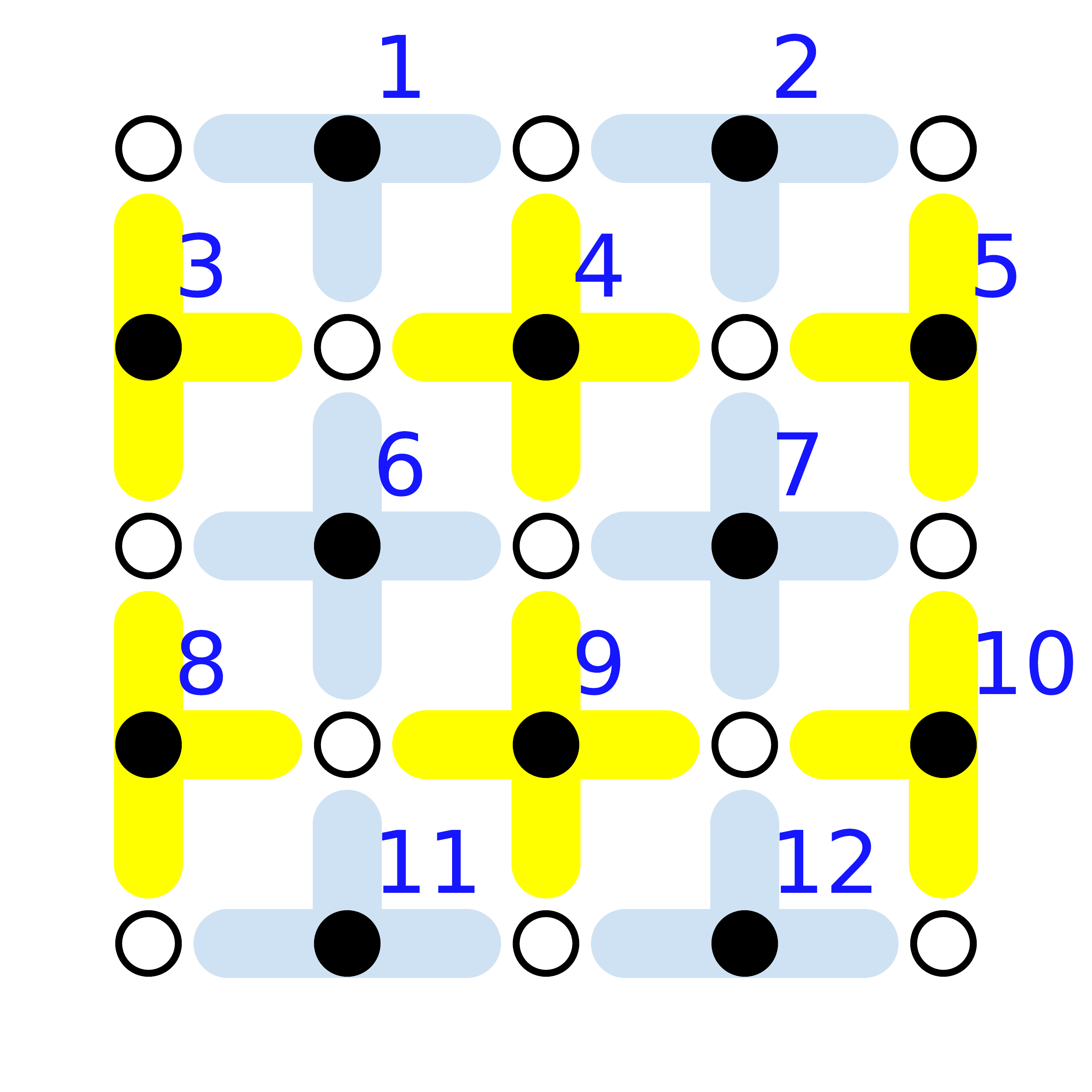}
        \caption{}
        \label{fig:graph_making_a}
    \end{subfigure}
    \begin{subfigure}{.24\linewidth}
        \centering
        \includegraphics[width=0.9\linewidth]{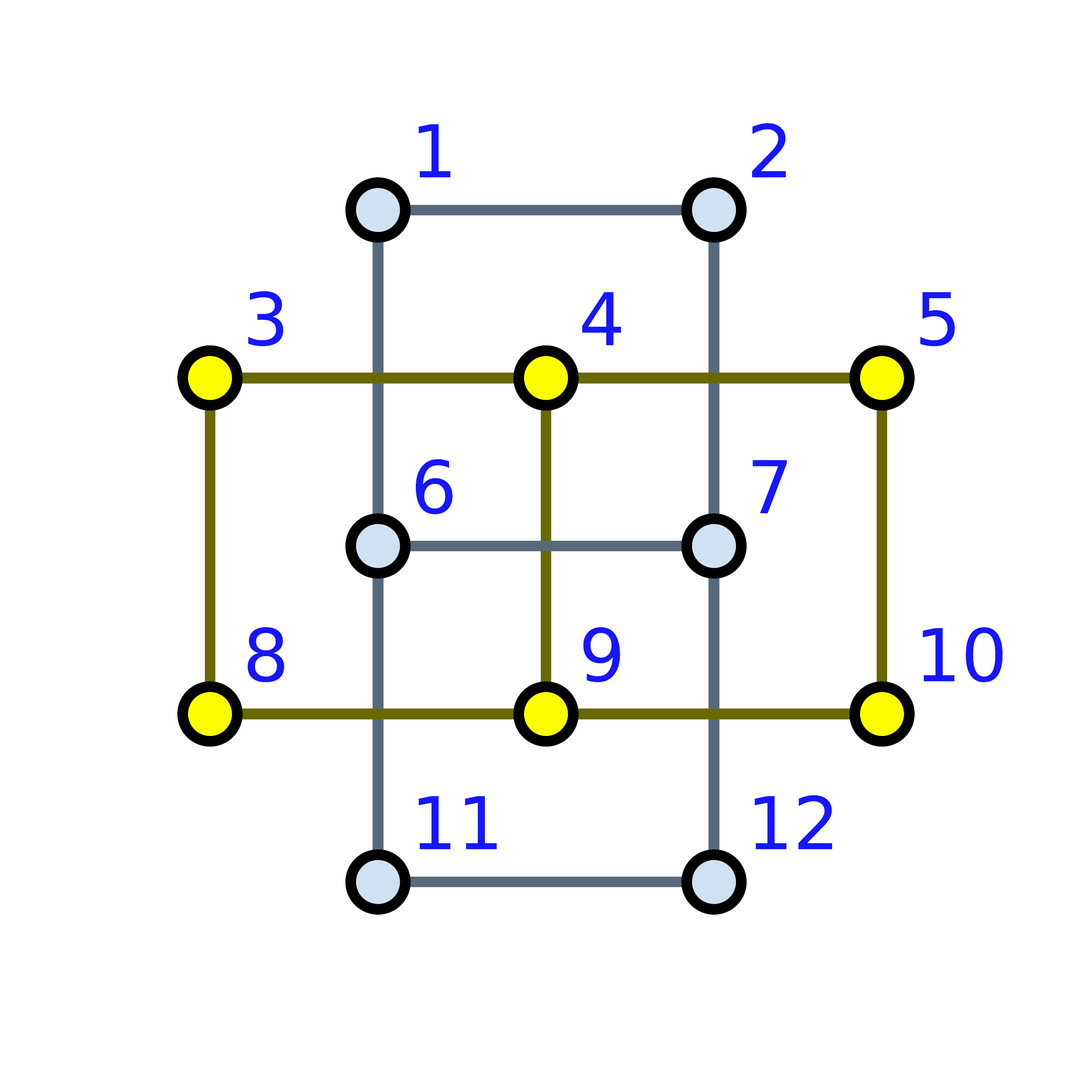}
        \caption{}
        \label{fig:graph_making_b}
    \end{subfigure}
    \begin{subfigure}{.24\linewidth}
        \centering
        \includegraphics[width=0.9\linewidth]{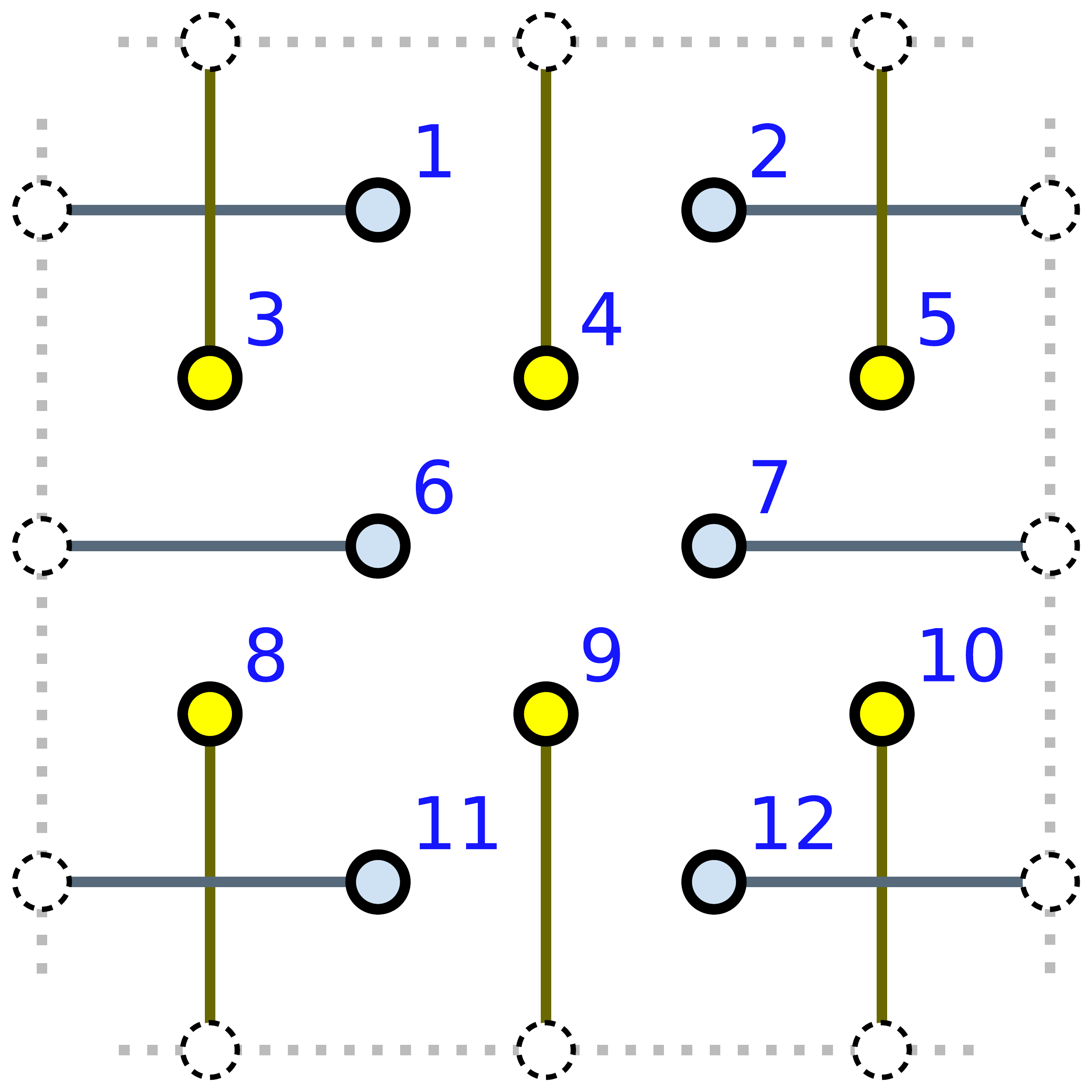}
        \caption{}
        \label{fig:graph_making_c}
    \end{subfigure}
    \begin{subfigure}{.24\linewidth}
        \centering
        \includegraphics[width=0.9\linewidth]{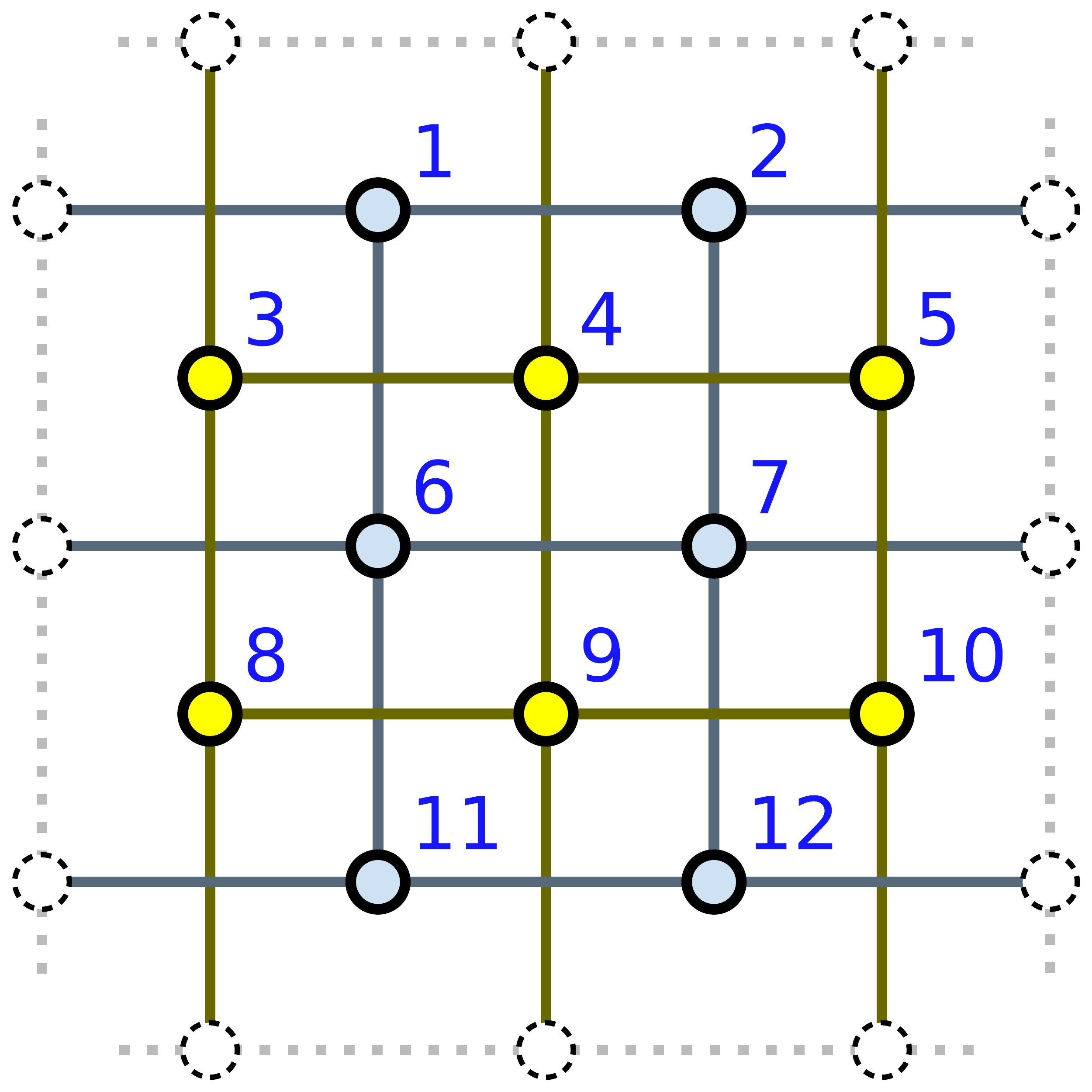}
        \caption{}
        \label{fig:graph_making_d}
    \end{subfigure}
    \caption{The process of generating a model graph for the graph for d = 3 CSS surface code. (a) The surface code patch with ancilla qubits numbered. (b) A graph showing all independent X and Z errors having two nontrivial measurements (c) A graph showing all independent X and Z errors having single nontrivial measurement (d) The model graph showing all independent X and Z errors. The dotted circles in (c) and (d) are virtual boundary vertices. In (b) (c) and (d), X and Z errors are shown blue and yellow respectively.
	}
    \label{fig:graph_making}
\end{figure*}

A vertex in the model graph represents an ancilla in the code.
An edge between two vertices represents an independent error (\eg X or Z error) of the corresponding data qubit that could result in nontrivial measurements of the corresponding ancillas.
For example,  \autoref{fig:graph_making_b} shows those edges in the model graphs for X and Z errors of CSS surface code.

An error on a data qubit at the border of the code may result in nontrivial measurement of a single ancilla, such as \autoref{fig:single_X_boundary_error}.
One can add a virtual \emph{boundary vertex} and use the edge connecting it with the ancilla vertex to represent this error. 
We call the boundary vertices on the left and right the X boundary vertices because they deal with X errors; similarly, we call the top and bottom boundary vertices the Z boundary vertices.
\autoref{fig:graph_making_c} visualizes these edges that connect one real vertex and one virtual boundary vertex.
The final model graph for the surface code of \autoref{fig:graph_making_a} is shown in \autoref{fig:graph_making_d}.

The weight of an edge in a model graph represents the probability of the corresponding independent error, i.e., $w_i = -\log\frac{P(\mathcal{E}_i)}{1 - P(\mathcal{E}_i)}$.
A set of edges corresponds to a set of independent errors; the probability of this set of errors is determined by the sum of the corresponding weights, per the independence assumption (\autoref{eq:independence}).

\begin{figure}[ht]
    \renewcommand*\thesubfigure{(\alph{subfigure})}  
    \centering
    \begin{subfigure}{.42\linewidth}
        \centering
        \includegraphics[width=1\linewidth]{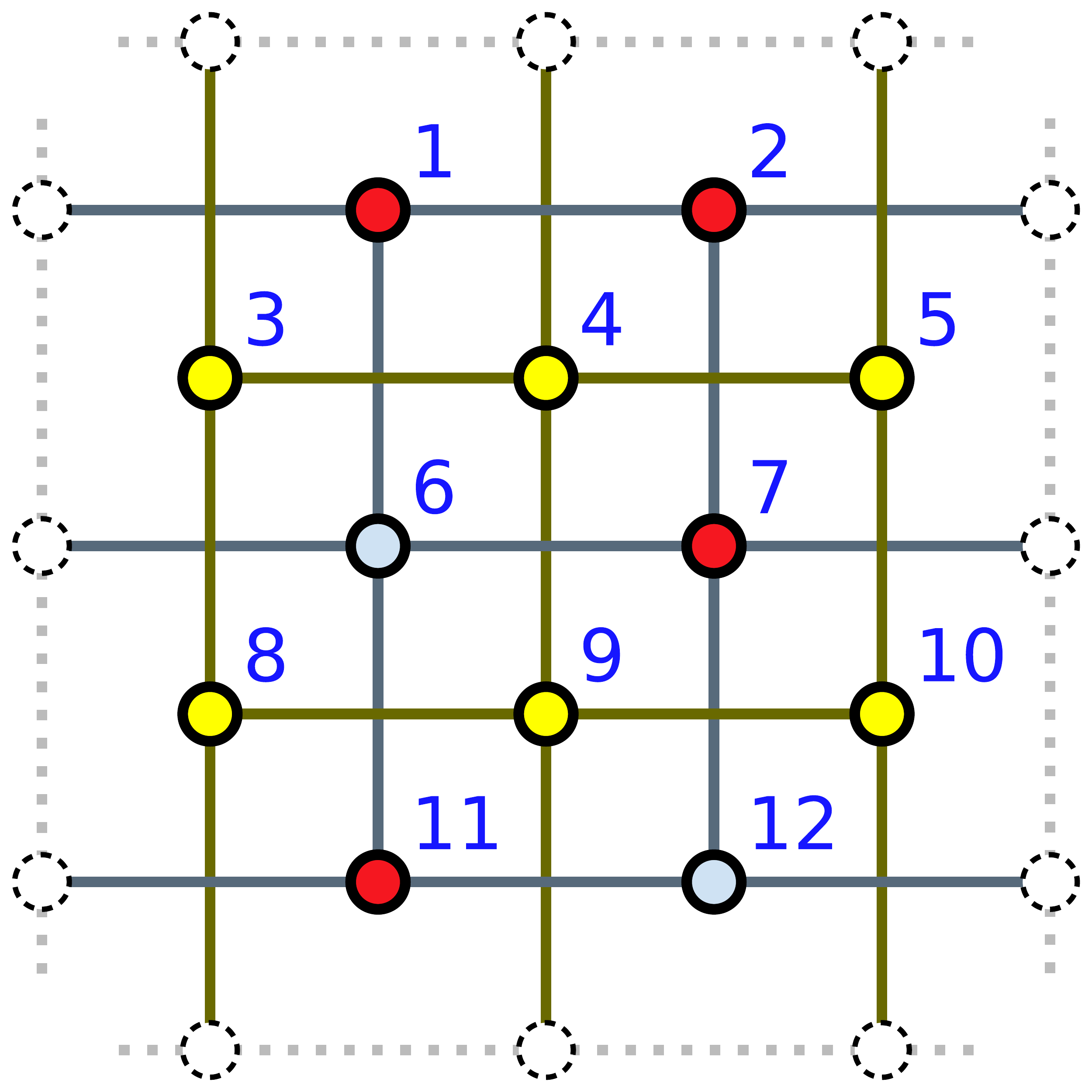}
        \caption{}
        \label{fig:csg_construction_c}
    \end{subfigure}
    \hfill
    \begin{subfigure}{.42\linewidth}
        \centering
        \includegraphics[width=1\linewidth]{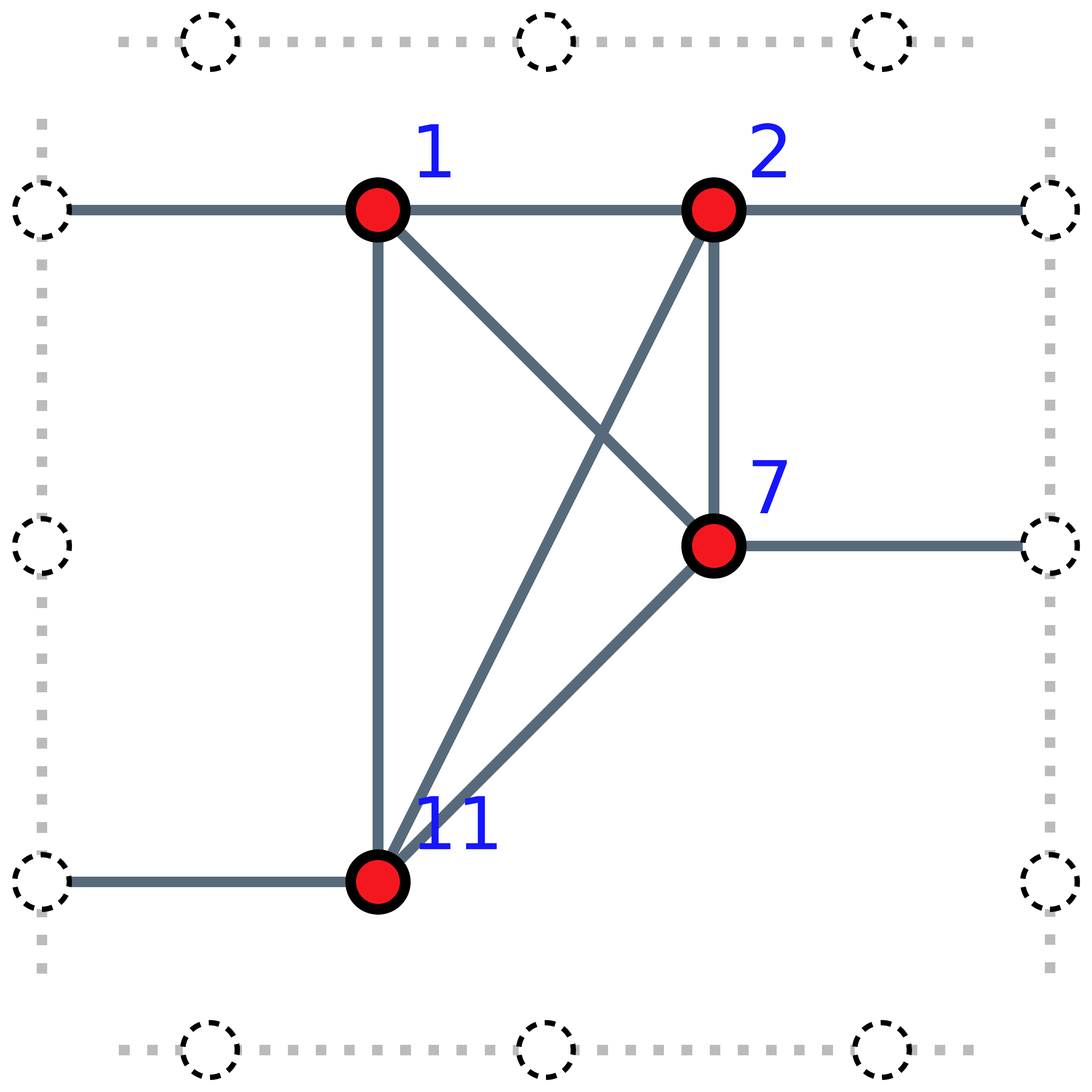}
        \caption{}
        \label{fig:csg_construction_d}
    \end{subfigure}
    \caption{ Example decoding (a) and syndrome (b) graphs for the model graph in \autoref{fig:graph_making_d}.
    The decoding graph is the model graph with with nontrivial measurement vertices marked (red). 
    The syndrome graph is constructed by computing the minimum-weight paths between nontrivial measurement vertices from the model graph and removing trivial ones.
    }
    \label{fig:csg_construction}
\end{figure}

\paragraph{Decoding Graph}
Once a measurement is performed on the surface code, some of the ancillas may report nontrivial measurement. 
The set of the corresponding vertices in the model graph is called the \emph{syndrome}.
The model graph and the syndrome are the basis of error decoding. Marking the vertices from the syndrome in the model graph, one derives the \emph{decoding graph}, which is the basis of error decoding. \autoref{fig:csg_construction_c} shows the decoding graph for the model graph from \autoref{fig:graph_making_d}.

\paragraph{Syndrome Graph}
Given the syndrome $S$,  
we can construct the \textit{syndrome graph} $G(V,E)$ as shown in \autoref{fig:csg_construction_d}. $V$ is the set of nontrivial measurement vertices from the model graph. $e\in E$ represents the set of the minimum-weight paths between two nontrivial measurement vertices in the model graph. Because each path in the model graph corresponds to an error pattern, $e$ also defines a set $\mathbf{e}$ of error patterns each corresponding to a minimum-weight path.

In a syndrome graph, we say two vertices are adjacent to each other if the edge connecting them has a small weight. Otherwise, we say they are far from each other. 

A subgraph of the syndrome graph is defined by a subset of $E$, $E'$. It defines a set of error patterns $\mathbf{E'}$ according to 
    $\mathbf{E}'=\{\E|\E=\sum_{e\in E'} \E_e,\forall \mathcal{E}_e\in\mathbf{e}\}$.

\paragraph{Logical Equivalence}
Given two subgraphs of the syndrome graph defined by $E_1$, $E_2\subseteq E$, respectively, we say they are logically equivalent if $\forall \mathcal{E}_1\in \mathbf{E}_1$ and $\forall \mathcal{E}_2\in \mathbf{E}_2$, $\mathcal{E}_1+\mathcal{E}_2$\footnote{Here $+$ is the symmetric difference, defined in \autoref{ap:proof}.} is a trivial logical operator.

\begin{table}[t]
\centering
\caption{Mathematical symbols}
\begin{tabular}{ |c|c|} 
 \hline
 Symbol & Meaning  \\ \hline\hline
 $e$ & edge\\\hline
 $E$ & set of edges  \\\hline
 $\mathcal{E}$ & an error pattern \\ 
 \hline
 $\hat{\mathcal{E}}$ & the operator form of $\E$ \\ 
 \hline
 
 $\mathbf{E}$& set of error patterns\\\hline
\end{tabular}
\end{table}

\subsection{Perfect Matching}
\label{sec:matching}

Given the syndrome $S$, the second stage seeks to identify the error pattern that caused it. That is, it solves the following problem, often approximately. \begin{equation}
    \mathcal{E}(S) = \arg \max \limits_\mathcal{E} P(\mathcal{E}|S)
    \label{eq:graphdecodeformula}
\end{equation}

Once $\mathcal{E}(S)$ is identified, a correction $C'$ can be applied,  to avoid a logical error, i.e., $ C'(S)\in \mathbf{C}_{\mathcal{E}(S)}$, or to perfectly cancel $\mathcal{E}(S)$, i.e., $\hat{C}'=\hat{\mathcal{E}}(S)^{\dagger}$.

Graph-based decoding solves \autoref{eq:graphdecodeformula} by finding a \emph{perfect matching} in the syndrome graph.
A perfect matching is a subgraph in which all vertices of the syndrome graph are incident to one and only one edge. It presents the set of error patterns $\mathbf{E}$ such that $\mathcal{E}\in\mathbf{E}$ produces the syndrome.
$P(\mathcal{E}|S)$ for $\mathcal{E}\in\mathbf{E}$ is determined by the sum of the weights of edges in the perfect matching. The larger the sum, the lower $P(\mathcal{E}|S)$.
Therefore, a most likely error pattern given the syndrome, \ie maximizing $P(\mathcal{E}|S)$, corresponds to a minimum weight perfect matching of the syndrome graph.

Some useful properties of perfect matchings of a syndrome graph can be found in \S\ref{sec:theory_pm}.

Comparing  \autoref{eq:graphdecodeformula} with \autoref{eq:coset}, one can see that graph-based decoding is missing the summation $\sum$. That is, it selects the most likely error and intends to correct it, while the optimal decoder should select the correction that will most likely correct the error. As a result, graph-based decoding is sub-optimal in terms of logical error rate. 

We note that virtual boundary vertices are treated specially when finding a perfect matching in the syndrome graph: they do not have to be matched. 

We care about two classes of graph-based decoders that differ in how they find a perfect matching given a syndrome graph. 

\begin{itemize}
\item 
 A \textit{Minimum Weight Perfect Matching (MWPM)} decoder finds a minimum-weight perfect matching for the syndrome graph. 
The best implementation of the MWPM decoder has a worst-case time complexity of $O(d^5)$~\cite{micali1980v,gabow1991faster,goldberg2004maximum}, not scalable for large surface codes. 

\item 
 A \textit{Union-Find (UF)} decoder~\cite{delfosse2020linear} does not find a minimum-weight perfect matching. It does not even find a perfect matching. Rather, it finds a subgraph that is logically equivalent to a perfect matching of low weight. 
 Because of this, UF decoders can be much faster and more scalable than MWPM decoders.
 Notably, the original UF decoder uses the decoding graph with $O(d^2)$ worst-case time complexity, instead of the syndrome graph (See \autoref{ap:equiv_uf}).
\end{itemize}

We will explain how these two classes are related further in \S\ref{sec:interpretation}.

\subsection{More Error Types}\label{ssec:generalize3d}
We now extend the graph-based decoding described above to deal with measurement errors and to approximate indirect errors, such as Pauli Y error.

\begin{figure}[!ht]
    \centering
    \includegraphics[width=0.8\linewidth]{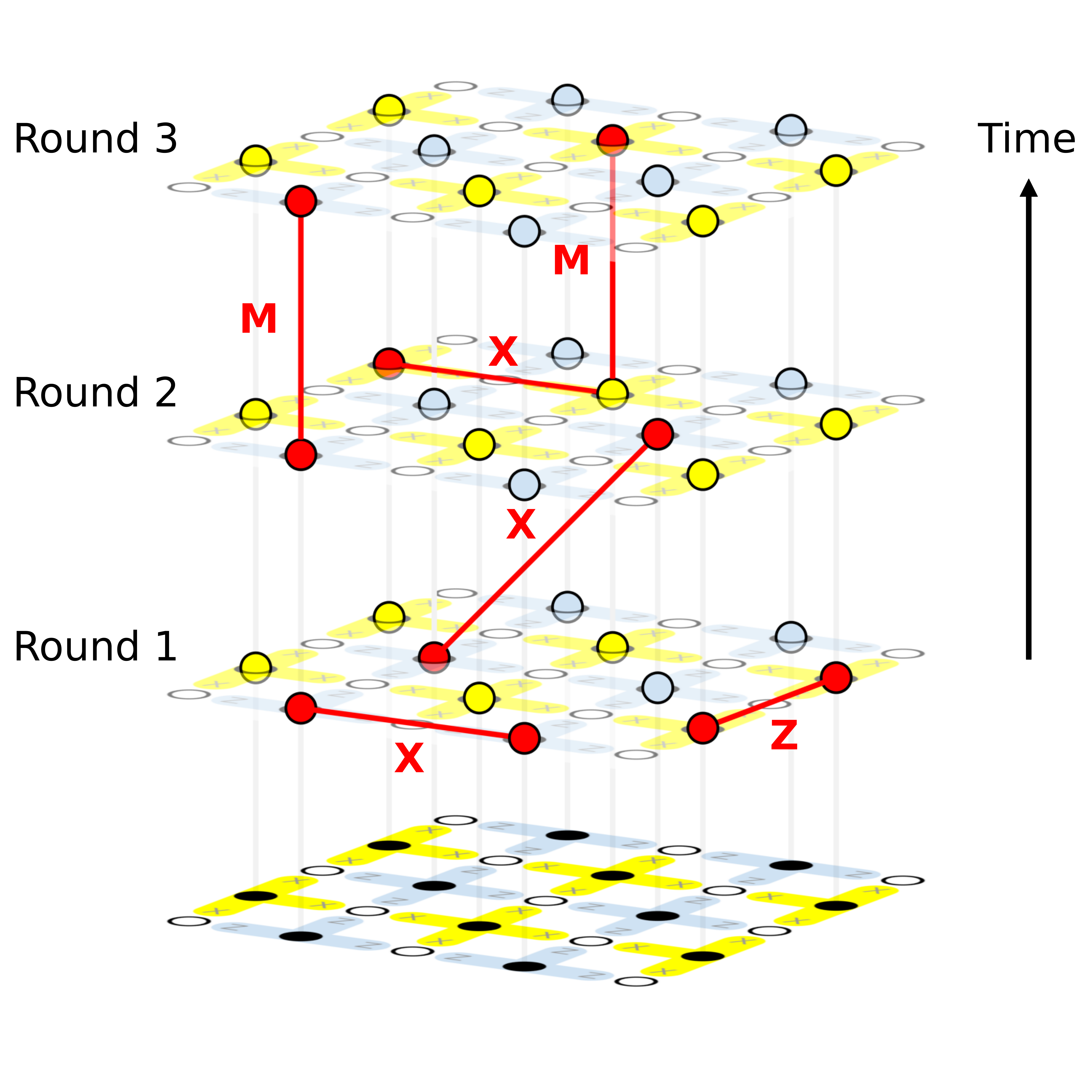}
    \caption{Decoding graph along with an error pattern on d = 3 surface code with 3 rounds of measurements. 
    This includes measurement errors and X or Z errors in ancillas propagating to data qubits. Measurement errors as well as errors propagating from ancilla to data qubits are represented by edges connecting adjacent layers (rounds). The error pattern shown has an isolated X error, isolated Z error, isolated measurement error (shown by M), an isolated X error in an ancilla propagated to data qubits (shown as a diagonal X edge ) and an error chain of X error and a measurement error. Error chains spanning multiple rounds make decoding even more complicated. }
    \label{fig:errors_3D}
\end{figure}

\textbf{Measurement errors}:~~Since ancilla qubits are error prone and the quantum gates to implement the stabilizer measurements are noisy, multiple rounds of ancilla measurements are required to perform decoding in a fault-tolerant manner.
Thus, the syndrome can be represented by a sparsely populated 3D tensor, where first layer indicate the ancilla values from the first round of measurement and subsequent layers indicate the difference of ancilla values in each round of measurement compared to the previous round of measurements.
The difference of ancilla values are chosen to ensure nontrivial measurements corresponding to each error is indicated only once in the 3D tensor.
Due to measurement errors, error chains can spread over multiple measurement rounds, thus increasing the space of total possible error patterns for a given syndrome as shown in \autoref{fig:errors_3D}.
This further complicates the decoding process, resulting in $2^{2d^3}$ different possible syndromes given $d$ noisy measurement rounds compared to $2^{2d^2}$ without these errors.

\textbf{Indirect errors} are errors that each impact more than two ancillas but can be approximated by multiple independent errors.
For example, Paul Y error in \autoref{fig:single_Y_error} can be decomposed as Pauli X error and Pauli Z error in \autoref{fig:single_X_error} and \autoref{fig:single_Z_error}, and thus can be decoded using the existing graph.
This decomposition implies that the probability of an indirect error is the product probabilities of its decomposed parts, which leads to inaccurate decoding results.
In the example above, if we assume the depolarizing error model where $P(X) = P(Y) = P(Z) = p$, the assigned probability $P'(Y) = P(X)P(Z) = p^2$ can be substantially smaller than the actual $P(Y)=p$ for small $p$.
As a result, graph-based decoding can only approximately solve the most likely error pattern in the existence of indirect errors.

\section{Interpretation of UF Decoder}
\label{sec:interpretation}

We next reveal the relationship between a union-find decoder~\cite{delfosse2021almost} and the famous blossom algorithm~\cite{edmonds_1965,kolmogorov2009blossom} that solves MWPM problems.

We will first introduce some necessary concepts in the blossom algorithm in \S\ref{sec:blossom} and then show how it solves the surface code decoding problem in \S\ref{ssec:blossom_sc}.
In \S\ref{sec:uf_interpretation}, we demonstrate that union-find decoders are close relatives of MWPM decoders.
Inspired by this, we describe a novel, general weighted union-find decoder in \S\ref{sec:weighted_uf}.

\subsection{Blossom Algorithm}
\label{sec:blossom}

The blossom algorithm uses linear programming (LP) to solve the MWPM problem~\cite{edmonds_1965,kolmogorov2009blossom} for a graph defined by $G(V, E)$ where $V$ is the set of vertices with even cardinality, and $E$ that of edges.
A matching $M$ of $G$ is a subset of $E$ in which no edges share a vertex.
A perfect matching $M_p$ is a matching whose elements cover all vertices of $G$. 
$\mathcal{O}$ is the set of subsets of $V$ with odd cardinaility, i.e. $\mathcal{O}=\{S| S\subset V; |S| \mathrm{~is~odd.}\}$.
For $e= ( u,v ) \in E$, if $u\in S$ and $v\notin S$, we say $S$ and $e$ are incident to each other.

In the primal problem of the linear programming formulation, a primal variable $x_e$ corresponds to $e \in E$. 
For each $S\in \mathcal{O}$, there is a \textit{primal constraint}. If $|S|=1$, exactly one edge incident to $S$ is in the solution. If $|S|>1$, at least one edge incident to $S$ is in the solution.
While $x_e$ is non-negative real, the primal constraints ensure that in one optimal solution,  $x_e$ is either $1$ or $0$ $\forall e \in E$,  representing a \emph{perfect matching}: $x_e = 1$ if $e$ is in the solution.
The \textit{primal objective function} $\sum_{e} w_e x_e$ is the total weight of the edges in the solution, to be minimized.

In the dual of the above problem~\cite[p. 81]{matousek2006understanding},
a \textit{dual variable} $y_S$ is defined for $S\in \mathcal{O}$, corresponding to a primal constraint.
Each dual constraint corresponds to an edge $e\in E$ (and its primal variable $w_e$): 
$\sum_{S\in\delta(e)} y_S \leq w_e$ where $\delta(e)$ = $\{S|S\in\mathcal{O}; S \mathrm{~incident~to~} e\}$.
The dual objective is to maximize $\sum_{S\in\mathcal{O}} y_{S}$.

The blossom algorithm leverages two insights, both based on the complementary slackness relationship between the primal and dual problems~\cite[p. 204]{matousek2006understanding}.
First, if $x_e>0$, i.e., $e$ is selected in the matching solution, the corresponding dual constraint must be \emph{tight}, i.e., $\sum_{S \in \delta(e)} y_S = w_e$. That is, the solution to the primal problem can only consist of \emph{tight} edges. 
Second, if $y_S>0$ for $S\in\mathcal{O}$ and $|S|>1$, the corresponding primal constraint must be tight: exactly one edge incident to $S$ is in the solution.

The second insight allows the algorithm to treat $S\in\mathcal{O}$ and $|S|>1$ like a vertex when $y_S>0$. Such $S$ are the eponymous \emph{blossoms}. Therefore, we use ``vertex'' to refer to both ordinary vertex and blossom below.

The first insight allows the algorithm to work on the primal and dual problems in an interleaving manner. 
When it works on the primal problem, it only considers the tight edges as candidates for the matching solution.
It identifies blossoms, odd number ($>1$) of ``vertices'' connected by tight edges in a circle, and alternating trees, odd number of ``vertices'' connected by tight edges in a tree where all leaves and nodes with multiple children are connected to the root through an even number of tight edges.
When it works on the dual problem, it adjusts $y_S$ to grow the dual objective function $\sum y_S$ while maintaining the tightness of edges in a blossom or alternating tree.
In doing so, it turns a dual constraint tight, that is $\sum_{S \in \delta(e)} y_S = w_e$, resulting in a new tight edge $e$.
With this new tight edge, the algorithm switches to work on the primal problem.   

\subsection{Blossom for Decoding Surface Code}
\label{ssec:blossom_sc}
When applying the blossom algorithm to a graph, 
we can imagine two vertices are separated by a distance of $w_e$, the weight of the edge incident to them. 
We can imagine that a \region covers a ``vertex'' $v$; 
for each edge incident to $v$, the \region covers  it by $y_v$, which is the dual variable.
The dual constraints dictate that the \regions would never overlap. When two \regions $u$ and $v$ meet on an edge $e=(u,v)$, $e$ becomes tight, i.e., $w_e=y_u+y_v$. 
We could also imagine a cluster, called \emph{\cluster}, started with a single \region.
When two \regions each from a different \cluster touch, the two \clusters merge.
Like \regions, \clusters do not overlap (i.e., share any vertices). A \cluster includes multiple \regions that touch each other in one way or another.
Because each \region $v$ corresponds to a non-zero dual variable $y_v$, we call the sum of the dual variables of the regions $\sum y_v$ within a \cluster the \csize of the \cluster.
We say a \cluster is even/odd if it contains an even/odd number of \regions.
If a \cluster covers virtual vertices from both left and right, we say it is \emph{attached}; 
otherwise, it is \emph{detached}. 
For example, in \autoref{fig:blossom-multi-grow-1.4} there are three \clusters, namely $\{A,B,C\}$, $\{D\}$, and $\{E\}$, all detached.

When the model graph is unweighted, this imagination can be conveniently visualized by the \emph{\diagram}~\cite{fowler2014minimum}.
In this diagram, the Manhattan distance between two vertices is the weight of the corresponding edge, i.e., $w_e$, as illustrated by \autoref{fig:blossom_multi_grow}.
We note that  the model graph of a surface code is unweighted if the data qubits have i.i.d. (independent and identically distributed) Pauli-X errors.

This imagination allows us to explain how the blossom algorithm works visually. Notably the algorithm actively operates on the internal of \cluster. 
When it works on the dual problem, it adjusts \regions to increase the \csize of each \cluster. For example, \autoref{fig:blossom-multi-grow-2.6} to \autoref{fig:blossom-multi-grow-2.9}.
Increasing the dual clusters may create a new tight edge and result in two \clusters merging into a ``larger'' one, in terms of the \csize and the number of vertices covered.
During the adjustment, the algorithm tries to keep tight edges within blossoms and alternating trees tight.

When the algorithm works on the primal problem, it only considers the tight edges. 
It will try to find a perfect matching inside each \cluster using only tight edges. When successful, it will mark that \cluster as \emph{solved} and will not work on it (and its \regions) unless the \cluster merges with another.
The algorithm may identify a blossom within a \cluster: an odd number of tight edges forming a circle, e.g., \autoref{fig:blossom-multi-grow-1.4}.
The algorithm then switches to work on the dual problem again by treating the blossom as a vertex and adjusting its own dual variable, e.g., $y_{\{A,B,C\}}$ in \autoref{fig:blossom-multi-grow-1.4} to \autoref{fig:blossom-multi-grow-2.9}, along with other dual variables.
The process repeats until all \clusters are solved.

By definition, a tight edge must be inside a \cluster because a tight edge corresponds to two touching \regions.
As a result, a matching solution to the primal problem must only include edges \emph{inside} \clusters. 

\subsection{Union-Find Decoder}
\label{sec:uf_interpretation}

We next show that a union-find (UF) decoder works on the syndrome graph to draw direct comparison with the blossom algorithm, using a similar visualization as used above. It is an adaptation from the original UF decoder~\cite{delfosse2021almost} that is based on the decoding graph. (See \autoref{ap:equiv_uf})

Unlike the blossom algorithm, which maintains the internal structure, i.e., \regions, for each \cluster, UF decoders only care about the number of vertices inside each cluster, i.e., whether a cluster is even or odd, and it only grows odd clusters.
In each step of growth, a UF decoder grows all odd clusters by the same amount: it makes sure the clusters do not overlap after growth. When two clusters touch, they get merged.
It stops growing a cluster when the cluster covers a virtual boundary vertex or becomes even. 
When no growth is possible, a UF decoder finds a subgraph as the solution for each cluster using only fully-grown edges.
This solution is logically equivalent to a perfect matching for the cluster that may use not-fully-grown edges.
Taken together, the solutions for all clusters form the solution for the syndrome graph, which is logically equivalent to a perfect matching. 
\autoref{fig:union_find_multi_grow} illustrates this procedure with the same \diagram from \autoref{fig:blossom_multi_grow}.
Different UF decoders may find different solutions within a cluster and as a result, may produce different solutions for the syndrome graph.

A UF decoder grows an odd cluster in the same way as the blossom algorithm grows a \cluster with a single ``vertex''.
It stops updating the even clusters while the blossom algorithm stops updating the solved \clusters for which a perfect matching with tight edges is found internally. 
Notably only a \cluster with an even number of ``vertices'' can be solved.

\subsection{Relationship between Blossom and UF}
\label{sec:relation}

At a high level, the blossom algorithm-based MWPM decoder and a UF decoder appear to be similar in that both decomposes the syndrome graph into non-overlapping subgraphs, i.e., clusters; both find the solution to the syndrome graph by aggregating the solutions of the subgraphs. 
Yet there are two key differences, which are behind the MWPM decoder's superiority in decoding accuracy and poor scalability. First, the MWPM decoder has a more sophisticated way to decompose the syndrome graph, or grow its \clusters. Second, it finds a minimum-weight perfect matching within a subgraph (\cluster) while the UF decoder is satisfied with finding a logical equivalent of a perfect matching. Intuitively, we have:

\textbf{Observation (UF/Blossom Similarity)}: given a syndrome graph, a UF decoder approximates the blossom-based MWPM decoder in accuracy, if the following two conditions are true.
\begin{itemize}
    \item Condition 1: They decompose the syndrome graph in a similar way. In the extreme case, there is a bijective mapping between their clusters such that the mapped clusters cover the same subset of vertices.
    
    \item Condition 2: Whether a perfect matching inside a cluster is minimum-weight does not matter.
\end{itemize}

We next examine situations when these two conditions may be true or close to be true. 
\subsubsection{Syndrome graph decomposed}
We first examine how they decompose the syndrome graph. While there are an enormous number of ways to decompose the graph, a number of factors constrain both the UF decoder and blossom algorithm so that they may end up decomposing the graph in a similar way. 

First of all, they have the same starting point: the same syndrome graph and the same set of clusters, each with a single vertex.

Second, when they terminate, their clusters must satisfy the following requirements: 
they must not overlap; each of them must have an even number of vertices because only even clusters can be solved; and a vertex is likely (but not always) to be in the same cluster as its nearest neighbor. The last is true because both grow all clusters by the same amount in each step of growth and merge clusters when they meet, forming literally clusters of vertices. 

Third, syndrome graphs that get decomposed into small (and therefore similar) clusters by both the UF decoder and blossom algorithm are more likely. This is due to the assumption that data qubit errors happen randomly and independently. As a result, nontrivial measurement outcomes are more likely to be randomly scattered in the model graph in pairs, resulting in a syndrome graph in which these pairs are much farther from each other than the two vertices within a pair. Such syndrome graphs are likely to get decomposed by both the UF decoder and blossom algorithm into small clusters each covering a pair or two.

Condition 1 also provides insight into why some revisions of UF decoders improve their accuracy~\cite{delfosse2021almost,huang2020fault}: these revisions allow a UF decoder to decompose a syndrome graph in a way closer to that the blossom algorithm would do.

\vspace{1ex}\textbf{Cluster vs. subgraph}:~~Given a syndrome graph, clusters (and \clusters) are defined by the vertices they cover while a subgraph is defined by the edges it includes.
Given a cluster, one can uniquely construct a subgraph, using the edges connecting any pair of vertices covered by the cluster. Likewise, given a subgraph, one can uniquely construct a cluster, using the vertices incidental to edges from the subgraph. Therefore, we use cluster and subgraph in an inter-exchangeable manner below, unless otherwise indicated.
For example, when we say a perfect matching for a cluster or inside a cluster, we are talking about the perfect matching for its corresponding subgraph. 
Interesting, if two clusters do not overlap, i.e., not sharing any vertex, their subgraphs do not overlap either, i.e., not sharing any edge.

\subsubsection{Equivalent matchings}
Assume the syndrome graph has been decomposed into non-overlapping clusters each with an even number of vertices.  Two perfect matchings $P_1$ and  $P_2$ for the syndrome graph are found by finding perfect matchings inside each cluster.

We know that a logical error happens when a chain of error connects left and right virtual vertices~\cite{bravyi1998quantum}, forming a nontrivial logical operator. 
Since a detached cluster cannot have such a chain in itself by definition, a subgraph of the cluster cannot represent a nontrivial logical operator. Therefore, we have the following Lemma and Theorem. See \autoref{ap:proof} for their proofs.

\noindent \textbf{Lemma (Equivalent Matchings)} if a cluster is detached, its perfect matchings are logically equivalent.
\newline

\noindent \textbf{Theorem (Equivalent Matchings)} if $P_1$ and $P_2$ are different only inside detached clusters, they are logically equivalent. \newline

When some \clusters are attached, the above cluster-based method will lead to a higher logical error rate than the MWPM decoders.
On the other hand, because usually most \clusters are detached, the difference in logical error rate can be small. Attached clusters are exponentially less likely with increasing code distance ($d$).

\section{Examples}
\label{sec:example}
We next provide two examples in which the two conditions described above are true and as a result, a UF decoder will be as accurate as an MWPM decoder.
\subsection{No adjacent errors}
When no errors are present in two adjacent data qubits, both conditions of the Observation are true and a UF decoder will achieve the same accuracy as MWPM decoders.
This is because when no errors are present in two adjacent data qubits, 
 vertices in the syndrome graph always appear in pairs that are far from each other or a single vertex adjacent to a virtual boundary vertex.
As a result, both a UF decoder and the blossom algorithm will decompose the syndrome graph into clusters each covering such a pair.

\subsection{XZZX code}
The XZZX surface code is a variant of the CSS surface code~\cite{bonilla2021xzzx}.
It employs a single type of stabilizer that measures $X\otimes Z\otimes Z\otimes X$.
Thus in the XZZX surface code, a single $X$ (or $Z$) error always generates a pair of nontrivial measurements horizontally (or vertically).

We next show the two conditions presented in \S\ref{sec:interpretation} are true for the XZZX code with infinite noise bias and noiseless stabilizers, known as the code capacity noise model~\cite{landahl2011fault}. We note the two conditions are not true with the circuit-level noise model~\cite{darmawan2021practical}.

Without loss of generality, we assume there are only Z errors on data qubits.
The model graph now becomes a set of disjoint subgraphs: each is a line (or 1D chain). As a result, the syndrome graph also becomes a set of parallel lines.
A perfect matching of the syndrome graph must comprise of perfect matchings for all such lines, each found separately.

\paragraph{Identical Clusters.}
An MWPM decoder grows an odd dual cluster at the same pace due to the use of the multiple tree approach~\cite{kolmogorov2009blossom}. When the dual cluster is 1D, it grows left and right at the same pace, exactly like how an odd cluster grows in a UF decoder.
In general, an MWPM decoder may not find a perfect matching with only tight edges inside an even dual cluster. 
However, inside a 1D even dual cluster, an edge between two adjacent vertices must be tight because of how these two vertices got merged into one cluster. As a result, an MWPM decoder can always find a perfect matching with tight edges by selecting edges between pairs of adjacent vertices. That is, in a 1D chain, an MWPM decoder always consider an even dual cluster solved, exactly like how a UF decoder treats an even cluster. 
Thus, the MWPM and UF decoders update clusters in exactly the same way and terminate at the same clusters when the clusters are 1D.

\paragraph{Equivalent Matchings}
When a cluster is detached, the solutions from the UF decoder and the MWPM decoder never differ by a nontrivial logical operator, according to \hyperref[lemma:equivalent_matchings]{\textit{Lemma (Equivalent Matchings)}}.
Otherwise, if the cluster is attached,  there are only two complementary perfect matchings of the cluster. Because a cluster grows left and right at the same rate, these two complementary perfect matchings must have the same weight and therefore both are MWPM. Because a UF decoder must select one of them, it will select a MWPM for the cluster and as a result, behave the same as an MWPM decoder for this cluster.

\section{Weighted Union-Find Decoder}
\label{sec:weighted_uf}
An astute reader will point out that the original union-find decoder~\cite{delfosse2021almost} works on the decoding graph. 
Yet the interpretation provided above works on the syndrome graph. This difference is only cosmetic as we purposefully adapt the union-find decoder for the syndrome graph in order to juxtapose it with the blossom algorithm. Implementation-wise, the decoding graph is preferred for lower time complexity. See \autoref{ap:equiv_uf} for more explanation.

Another, more substantial difference is that the original UF decoder works on an \emph{unweighted} decoding graph, assuming identical error probability for all data qubits. 
As a result, it grows all odd clusters by half a unit each step.
Yet the interpretation above does not need this assumption. Rather, it must compute the safe amount of growth for the odd clusters at each step.
This leads to a more general union-find decoder that works with weighted model graphs, described below.

Huang, Newman and Brown~\cite{huang2020fault} already report a UF decoder design that uses weighted model graphs, without explicitly identifying the link between the UF and MWPM decoders presented above. They compute the weight as $\ln((1-p)/p)$ where $p$ is the error probability, which is similar to the integer-weighted UF decoder described below. 

\paragraph{Real-Weighted Union-Find Decoder}
A real-weighted union-find decoder has a time complexity no worse than $O(N^2)$, $N$ being the number of vertices in the model graph.
This is due to two factors: first, in each step, it has a time-complexity of at most $O(N)$ to compute the maximum safe growth such that when all odd clusters grow by that much, they will not overlap.
The maximum safe growth is calculated such that at least one more edge is fully covered by clusters, i.e., becomes fully grown, to use the language of Delfosse and Nickerson~\cite{delfosse2021almost}.
Second, it takes at most $O(N)$ steps to finish grow all clusters because each step will get at least one edge covered by the clusters while there are $O(N)$ uncovered edges to begin with.

A real-weighted union-find decoder should have  better time complexity, worst-case and average, than the blossom algorithm, for three reasons.
First, the blossom algorithm works on the syndrome graph, which has $O(N^2)$ edges, while the union-find decoder works on the model graph, which has $O(N)$ edges.
Second, the \clusters in the blossom algorithm may shrink, while the clusters in the union-find decoder only grow. The possibility of shrinking may lead to more steps before finish growing all clusters.
Finally, the blossom algorithm maintains the structure inside each cluster. As a result, it merges two clusters with a time complexity proportional to the cluster size, while the real-weighted UF decoder merges two clusters within constant time.

\paragraph{Integer-Weighted Union-Find Decoder}
When the weights of the model graph are small integers, union-find decoders can be much faster in terms of the worst-case time complexity.
This is because the safe growth computation becomes trivial: 
all odd clusters grow by half a unit each step, just like in the original union-find decoder.
Each edge with weight $w_e$ in the model graph is at most visited $O(w_e)$ times.
Given the maximum weight $w_{\max}$, the overall worst-case time complexity of integer-weighted union-find decoder is $O(N \cdot (\alpha(N) + w_{\max}))$ where $\alpha(N)$ is an almost constant inverse Ackermann’s function~\cite{delfosse2021almost}.
As a result, the average decoding time complexity must be between $O(N)$ and $O(N \cdot (\alpha(N) + w_{\max}))$, which is almost linear.

On the other hand, we can see the worst time complexity of an integer-weighted UF decoder grows with $w_{\max}$ while that of a real-weighted union-find decoder does not. 
Therefore, when $w_{\max}$ is large, a real-weighted union-find decoder can be faster than an integer-weighted one when the integer weights are too large.

\paragraph{Implementation}
We have implemented and open-sourced both weighted UF decoders described above~\cite{qec-playground}.
Our implementation of the real-weighted UF decoder uses the integer data type (64-bit signed integer) to avoid rounding errors in the floating point data type, a strategy borrowed from the blossom V algorithm implementation~\cite{kolmogorov2009blossom}.
For both weighted UF decoders, we use an integer $w_{\max}$ to represent the largest weight $W = \max_e{w_e}$ and compute the integer weight for an edge of weight $w_e$ as $\lfloor w_e /W * w_{\max}\rfloor$.
The only difference between our implementations of the real-weighted and integer-weighted UF decoders lies in how they compute the growth when growing the odd clusters: the real-weighted UF decoder computes the maximum safe growth while the integer-weighted one grows them by one each time, as explained in~\S\ref{sec:weighted_uf}.
As a result, they have the same accuracy given the same scaling $w_{\max}$.

\section{Discussion}
\label{sec:discussion}

In this paper, we limit the discussion to open-boundary surface codes. Nevertheless, we believe the same interpretation works on other surface codes as well with proper generalization of notions.
For example, ``attached cluster'' can be generalized to a subgraph whose edges constitute a nontrivial logical operator.
The relationship between UF decoders and the blossom algorithm also suggests that UF decoders can be adapted to solve any decoding problem that can be solved by a MWPM of the syndrome graph, e.g., the color code ~\cite{sahay2022decoder}.

The revealed relationship between UF decoder and the blossom algorithm further suggests ways of cross-pollination between the blossom algorithm and UF decoders.
We have already showed that it can lead to new UF decoder designs in \S\ref{sec:weighted_uf}. 
One can borrow more ideas from the blossom algorithm to make UF decoders better. 
For example, one can keep the internal structures of small clusters so that they grow like the \clusters in the blossom algorithm.
On the other hand, one could also bring ideas from UF decoders into the blossom algorithm to improve the latter's speed.
For example, we have recently shown that instead of the syndrome graph, MWPM decoders can be made faster~\cite{fusion-blossom} by adopting the decoding graph used by UF decoders, which has also been independently discovered by Higgott and Gidney~\cite{pymatchingv2}.

\bibliographystyle{naturemag}
\bibliography{ref}{}

\section*{Acknowledgements}
We thank Shruti Puri for fruitful discussion and insightful feedback. This work was supported in part by Yale University and NSF MRI Award \#2216030.

\newpage
\appendix
\begin{center}
    \textbf{\Large Appendix}
\end{center}

\section{Surface Code, Syndrome Graph, and Perfect Matching}
\label{ap:proof}

\subsection{Structured Proof}\label{ap:structured-proof}
We opt for the style of proof known as structured ~\cite{lamport2012write} for its clarity.
A structured proof employs a hierarchical structure that provides both the high-level idea and low-level details.
The proof consists of steps organized in multiple layers.
Each step is prefixed with $\langle n \rangle m$, which means it is the $m$-th step of the $n$-th layer.
Every step comes with a proof, either a single sentence prefixed with \proofPfkwd\ or a sub layer ending with \qedstepPfkwd.

\subsection{Error Pattern, Syndrome, and Parity}
A surface code can be represented by the model graph $G(U,F)$, where $U$ is the set of vertices, each representing a stabilizer, and $F$ is the set of edges, each representing an independent error source, including data qubit.
$\forall u\in U$, let $F(u)$ denote the set of edges that are incident to $u$; likewise, $\forall f\in F$, $U(f)$ denotes the set of vertices that are incident to $f$.

\paragraph{Error Pattern}
In this work, we only consider Pauli errors. Therefore, an error to $f\in F$ can be regarded as an operator $\hat{f}$ that ``flips'' the state of $f$. Obviously $\hat{f}\hat{f}=I$.
An error pattern $\E$ is a subset of error sources (edges) in which error happens. On the model graph, it can be denoted as $E\subseteq F$. 
An error pattern can also be considered as an operator $\hat{\E}$ that flips the corresponding error sources: $\hat{\E}=\prod_{f\in E} \hat{f}$. 
Because there is a bijective mapping between error patterns and operators, we often use an error pattern $\E$ and its operator $\hat{\E}$ in an interchangeable manner.

Borrowing the notations from~\cite{bombin2013introduction}, we can represent $\E$ as a sum over all edges: $\E=\sum_{f\in E}f$.

With this sum form, two error patterns can be added together to form a new one with $f+f=0$, because an error flips the state of error source $f$.
Given two error patterns, $\E_1$ and $\E_2$, we define their sum as 
\[ \E_1\oplus\E_2= \sum_{f\in E_1}f+\sum_{f\in E_2}f=\sum_{f\in E_1 \triangle E_2}f\]

The error patterns and $\oplus$ constitute an Abelian group. The inverse of an error pattern is simply itself. For simplicity, we use $+$ in place of $\oplus$.

A special type of error pattern is a path of contiguous edges that starts with one vertex and ends with another. A circle is a path that starts and ends with the same vertex.

\paragraph{Syndrome}
For a stabilizer $u\in U$, it gives a nontrivial measurement outcome if an odd number of edges it connects with experience error.  Intuitively, one can consider an error in an edge incidental to $u$ ``flips'' the stabilizer measurement outcome.

For error patterns of a single error, $\forall f\in F$, we can represent its syndrome in a similar sum form  
\[S(\{f\})=\sum_{u\in U(f)} u.\] 
 That is, a single error in $f$ causes nontrivial measurement outcomes in all stabilizers it is connected with. 

Given an error pattern $\E$, its syndrome $S$ includes all the nontrivial measurement outcomes. We can represent $S$ as a sum over the syndromes of all its errors.
\begin{equation}
S(\E)=\sum_{f\in E}S(\{f\})
\end{equation}
where $\forall u\in U$, $u+u=0$, because a stabilizer gives a nontrivial measurement outcome if an odd number of edges it connects with experience error. 

Given two error patterns $\E_1$ and $\E_2$, we have 

\[S(\E_1+\E_2)=S(\E_1)+S(\E_2).\]

Given a path $\mathcal{P}$ in the model graph connecting two vertices $u$ and $v$, we have $S(\mathcal{P})=u+v$.

\paragraph{Parity}
For $u\in U$, let $V_L(u)$ indicate whether $u$ is a left virtual boundary vertex: 1 if yes; 0 if not. 

For error patterns of a single error, $\forall f\in F$, its left parity $P_L(\{f\})$ is defined as parity of how many left virtual boundary vertices $f$ is incident to. 
\[P_L(\{f\})=\sum_{u\in U(f)} V_L(u)\]
where $V_L(u)+V_L(u)=0$. Given an error pattern $\E$, its left parity $P_L$ is the sum of the parity of all its single errors. 
\begin{equation}
P_L(\E) = \sum_{f\in E} P_L(\{f\}).
\end{equation}
where $1+1=0$. Given two error patterns $\E_1$ and $\E_2$, we have 

\[P_L(\E_1+\E_2) = P_L(\E_1)+P_L(\E_2).\]

Given a path $\mathcal{P}$ in the model graph connecting two vertices $u$ and $v$, we have $P_L(\mathcal{P})=V_L(u)+V_L(v)$.

The parity on the right boundary $P_R(\E)$ is similarly defined. For open-boundary surface code, we have: 
\begin{equation}
\label{equ:parity-syndrome}
P_L(\E) + P_R(\E) = |S(\E)| \ (\mathrm{mod}\ 2).
\end{equation}

We provide some intuition behind this property that links the parity with the syndrome. 
When the error pattern is empty, $P_L(\E) + P_R(\E) = |S(\E)| \ (\mathrm{mod}\ 2) = 0$.
Otherwise, $\forall f=\langle u,v \rangle \in E$, if $V_L(u)=1$, i.e., $f$ is incidental to the left virtual boundary, $v$ must be a stabilizer. We have $P_L(\{f\})=1$ and $S(\{f\})=v$. That is, $f$ will flip both $P_L(\E)$ and $|S(\E)| (\mathrm{mod}\ 2)$. The same can be said about edges incidental to the right virtual boundary.
 if $f$ is not incidental to either virtual boundary, $P_L(\{f\})=P_R(\{f\})$=0 and $f$ will flip the states of both $u$ and $v$ and as a result, leave $|S(\E)| (\mathrm{mod}\ 2)$ unchanged.

\paragraph{Logical operator and trivial logical operator} 
When an operator is applied to the surface code, a subset of the stabilizers may have nontrivial measurement outcomes. When the subset is empty, i.e., all measurement outcomes are trivial, the operator is a \emph{logical operator}. That is, a logical operator does not produce any syndrome. 

\vspace{1ex}\noindent \emph{Definition (Logical Operator)}: an error pattern $\E$ is a logical operator if and only $S(\E)=0$.

If $\E$ is a logical operator,  
we have $S(\E) = 0$ and therefore $P_L(\E) = P_R(\E)$. Therefore, when we know $\E$ is a logical operator, we will use $P(\E)$ as a short-hand for both $P_L(\E)$ and $P_R(\E)$.

A trivial logical operator does not change the logical state of the surface code. 
In \S\ref{sec:background_surface}, we introduced the notion of trivial logical operator and mentioned that error patterns that form closed circles are trivial logical operators. We define it formally below.

\vspace{1ex}\noindent \emph{Definition (Trivial Logical Operator)}:~~A logical operator $\E$ is trivial iff $P(\E)=0$.

A circle on the model graph is a trivial logical operator. This is because a circle $\mathcal{C}$ starts and ends with the same vertex $u$: $S(\mathcal{C})=u+u=0$ and $P(\mathcal{C})=V_L(u)+V_L(u)=0$.

Given any trivial logical operator $\mathcal{T}$ and $\E$ an error pattern, we have $S(\E+\mathcal{T})=S(\E)$ and $P(\E+\mathcal{T})=P(\E)$. That is, adding a trivial logical operator to any error pattern will not change the syndrome or parity. 

\subsection{Syndrome Graph}
\label{sec:theory_syndromegraph}

Given the surface code, represented by the model graph $G(U,F)$, and its measurement, we can construct the syndrome graph $G(V,E)$ as follows. $V\subseteq U$ includes all the model graph vertices that have nontrivial measurement outcomes; $E=\{e=\langle u,v\rangle| \forall u, v\in V\}$. For $e=\langle u,v\rangle\in E$, the weight $w_e$ is computed as the weight of the minimum-weight path between $u$ and $v$ in the model graph. 

Because there may be multiple minimum-weight paths between $u$ and $v$, $e=\langle u,v\rangle$ may represent multiple error patterns, which is denoted by the set $\mathbf{e}$. $\forall\E\in\mathbf{e}$, it can be considered a collection of single qubit errors, corresponding to the model graph edges in a minimum-weight path $\mathcal{P}$. That is, $\E=\sum_{f\in \mathcal{P}} f$.

\vspace{2ex}\noindent \textbf{Lemma (O)}\label{lemma:pm0} $\forall e\in E$, $\mathcal{E}_1$, $\mathcal{E}_2 \in \mathbf{e}$, $\mathcal{E}_1+\mathcal{E}_2$ is a trivial logical operator.

\begin{proof2}
\suffices{ $S(\E_1+\E_2)=0$ and $P(\E_1+\E_2)=0$.}
    \begin{proof2}
    \pf\ By definition of trivial logical operator. 
    \end{proof2}
\step{1} {$\E_1$ and $\E_2$ are paths connecting the same two vertices $u$ and $v$ in the model graph.} 
    \begin{proof2}
    \pf\ By definition of $e$.
    \end{proof2}
\step{2}{$S(\E_1+\E_2)=0$}
    \begin{proof2}
    \step{2.1}{$S(\E_1)=u+v$; $S(\E_2)=u+v$}
    \step{2.2}{$S(\E_1+\E_2)=S(\E_1)+S(\E_2)$\\$= (u+u)+(v+v)=0$;}
    \qedstep
    \end{proof2}  
\step{3}{$P(\E_1+\E_2)=0$}
    \begin{proof2}
    \step{3.1}{$P(\E_1)=V_L(u)+V_L(v)$,\\ $P(\E_2)=V_L(u)+V_L(v)$}
    \step{3.2}{$P(\E_1+\E_2)=P(\E_1)+P(\E_2)=\\ (V_L(u)+V_L(u))+(V_L(v)+V_L(v))=0$}
    \qedstep
    \end{proof2}
\qedstep
\end{proof2}

\paragraph{Subgraph}
A subgraph of the syndrome graph $G(V,E)$ is defined by a subset of $E$, $E'\subseteq E$.
$E'$ defines a set of error patterns $\mathbf{E}'$.
\begin{center}
    $\mathbf{E}'=\{\E|\E=\sum_{e\in E'} \E_e,\forall \E_e\in\mathbf{e}\}$
\end{center}
Therefore, $\forall \E\in \mathbf{E}'$, $\exists \E_e\in \mathbf{e}$ for $\forall e\in E'$ such that $\E=\sum_{e\in E'} \E_e$. 
That is, an error pattern represented by the subgraph can be ``decomposed'' into error patterns represented by its edges. 

Using the familiar sum form, we can represent $E'$ as $\sum_{e\in E'}e$.
 Two subgraphs can be ``added'' together to form a new one with $e+e=0$. The symmetric difference between two subgraphs $E_1$ and $E_2$ can be simplified as $E_1\triangle E_2 = E_1+E_2$.

\subsection{Perfect matchings}
\label{sec:theory_pm}

Given a syndrome graph $G(V,E)$, a perfect matching is a subgraph in which every vertex from $V$ is incident to one and only one edge from $E$. 
It represents a set of error patterns $\mathbf{E}$ such that $\mathcal{E}\in\mathbf{E}$ produces the syndrome. That is, $S(\E)=V$.
We note that $\mathbf{E}$ does not include all the error patterns for the syndrome. 

\noindent \textbf{Lemma (I)}\label{lemma:pm1} Given a syndrome graph and its perfect matching represented by $\mathbf{E}$ and $\mathcal{E}_1$, $\mathcal{E}_2 \in \mathbf{E}$, $\mathcal{E}_1+\mathcal{E}_2$ is a trivial logical operator.

\begin{proof2}
\suffices{$S(\E_1+\E_2)=0$ and $P(\E_1+\E_2)=0$.}
\step{1}{$S(\E_1+\E_2)=0$}
    \begin{proof2}
    \step{1.1}{$S(\E_1)=S(\E_2)$=V}
        \begin{proof2}
        \pf\ By definition of perfect matching.
        \end{proof2}
    \step{1.2}{$S(\E_1)+S(\E_2)=V+V=0$}
    \qedstep
    \end{proof2}
\step{2}{$P(\E_1+\E_2)=0$}
    \begin{proof2}
    \step{2.1}{Let $E'\subseteq E$ denote the edges of the perfect matching}
    \step{2.2}{$\exists \E^1_e\in \mathbf{e}$ for $\forall e\in E'$ such that $\E_1=\sum_{e\in E'} \E^1_e$; $\exists \E^2_e\in \mathbf{e}$ for $\forall e\in E'$ such that $\E_2=\sum_{e\in E'} \E^2_e$.}
        \begin{proof2}
        \pf\ By subgraph decomposition.
        \end{proof2}
    \step{2.3}{$P(\E_1+\E_2)=P(\E_1)+P(\E_2)\\ =P(\sum_{e\in E'} \E^1_e)+P(\sum_{e\in E'} \E^2_e)\\ =\sum_{e\in E'}P(\E^1_e)+\sum_{e\in E'}P(\E^2_e)\\ =\sum_{e\in E'}(P(\E^1_e)+P(\E^2_e))\\ =\sum_{e\in E'}(P(\E^1_e+\E^2_e))=0$}
        \begin{proof2}
        \pf\ By \hyperref[lemma:pm0]{Lemma (O)}.
        \end{proof2}
    \end{proof2}
\qedstep
\end{proof2}

\noindent \textbf{Lemma (II)}\label{lemma:pm2} Given a syndrome graph, let $\mathbf{E}_1$ and $\mathbf{E}_2$ denote two perfect matchings.  For $\mathcal{E}_1 \in \mathbf{E}_1$, $\mathcal{E}_2 \in \mathbf{E}_2$, $\mathcal{E}_1+\mathcal{E}_2$ is a logical operator.

\begin{proof2}
\suffices{$S(\E_1+\E_2)=0$}
\step{1}{$S(\E_1)=S(\E_2)=V$}
    \begin{proof2}
    \pf by definition of perfect maching
    \end{proof2}
\step{2}{$S(\E_1+\E_2)=S(\E_1)+S(\E_2)=V+V=0$}
\qedstep
\end{proof2}

\subsubsection{Cluster decomposition}
\label{sec:theory_cluster}

Given a cluster defined by an even subset of $V$. The above lemmas are also true for perfect matchings inside the cluster.
We will refer to them as Lemma (Cluster) in the following.

\subsection{Equivalent Matchings}
\label{sec:ap_eq}

Assume the syndrome graph has been decomposed into non-overlapping clusters each with an even number of vertices $C_i$, $i=1,2,..., n$. 
Let $G_i$, $i=1,2,...,n$ denote the corresponding subgraphs.  $P_i^1$ and $P_i^2$ denote two perfect matchings for $G_i$. $P_1=\sum_i P_i^1$ and  $P_2=\sum_i P_i^2$ are two perfect matchings for the syndrome graph.

Let $\mathbf{E}_i^1$ and $\mathbf{E}_i^2$ denote sets of error patterns represented by  $P_i^1$ and $P_i^2$, respectively. Let $\mathbf{E}_1$ and $\mathbf{E}_2$ denote sets of error patterns represented by  $P_1$ and $P_2$, respectively.
$\forall \E_1\in \mathbf{E}_1$, $\exists \E_i^1\in\mathbf{E}_i^1$, such that $\E_1=\sum_i \E_i^1$. Similarly, $\forall \E_2\in \mathbf{E}_2$, $\exists \E_i^2\in\mathbf{E}_i^2$, such that $\E_2=\sum_i \E_i^2$.

\noindent \textbf{Lemma (Equivalent Matchings)}\label{lemma:equivalent_matchings} if a cluster is detached, its perfect matchings are logically equivalent.
\newline

\begin{proof2}
\suffices{$\forall \mathcal{E}_1\in \mathbf{E}_1$ and $\forall \mathcal{E}_2\in \mathbf{E}_2$, $\mathcal{E}_1 + \mathcal{E}_2$ is a trivial logical operator}
    \begin{proof2}
        By definition of \emph{Logical Equivalence} for subgraphs
    \end{proof2}
\step{1}{$S(\E_1+\E_2)=0$}
    \begin{proof2}
    \step{1.1}{$S(\E_1+\E_2)=S(\E_1)+S(\E_2)=0$}
        \begin{proof2}
        \pf\ $S(\E_1)=S(\E_2)$ because both perfect matchings produce the syndrome inside the same cluster.
        \end{proof2}
    \qedstep
    \end{proof2}  
\step{2}{$P(\E_1+\E_2)=0$}
    \begin{proof2}
    \step{2.1}{ Either $P_L(\E_1)=P_L(\E_2)=0$ or $P_R(\E_1)=P_R(\E_2)=0$}
        \begin{proof2}
        \pf\ by definition of \emph{Detached cluster}.
        \end{proof2}
    \step{2.2}{$P_R(\E_1+\E_2)=P_L(\E_1+\E_2)$}
        \begin{proof2}
        \pf\ by \autoref{equ:parity-syndrome} and Step \stepref{1}.
        \end{proof2}
    \qedstep
    \end{proof2}
\qedstep
\end{proof2}

\noindent \textbf{Theorem (Equivalent Matchings)} if $P_1$ and $P_2$ are different only inside detached clusters, they are logically equivalent. \newline

\begin{proof2}
\suffices{ $\forall \E_1\in \mathbf{E}_1$ and $\forall \E_2\in \mathbf{E}_2$, $S(\E_1+\E_2)=0$ and $P(\E_1+\E_2)=0$.}
    \begin{proof2}
    \pf\ By definition of trivial logical operator and definition of logical equivalence.
    \end{proof2} 
\step{1}{$S(\E_1+\E_2)=0$}
    \begin{proof2}
    \step{1.1}{$S(\E_1+\E_2)=S(\sum_i(\E_i^1+\E_i^2))\\ =\sum_i S(\E_i^1+\E_i^2)=0$}
    \begin{proof2}
        \pf\ {$S(\E_i^1+\E_i^2)=0$} by \hyperref[lemma:pm2]{Lemma (II)}.
    \end{proof2}
    \qedstep
    \end{proof2}  
\step{2}{$P(\E_1+\E_2)=0$}
    \begin{proof2}
    \step{2.1}{$P(\E_1+\E_2)=P(\sum_i(\E_i^1+\E_i^2))\\ =\sum_i P(\E_i^1+\E_i^2)=0$}
        \begin{proof2}
            \step{2.1.1}{If cluster $i$ is detached, $P(\E_i^1+\E_i^2)=0$ }
            \step{2.1.2}{Otherwise, cluster is attached. $P_i^1=P_i^2$ by assumption and then $P(\E_i^1+\E_i^2)=0$}
                \begin{proof2}
                \pf\ by \hyperref[lemma:pm2]{Lemma (II)} (Cluster).
                \end{proof2}
            \qedstep
        \end{proof2}
    \qedstep  
    \end{proof2}
\qedstep
\end{proof2}

\section{UF Decoder on Syndrome Graph}\label{ap:equiv_uf}

\begin{figure}[t]
    \renewcommand*\thesubfigure{(\arabic{subfigure})}  
    	\centering
	\begin{subfigure}{.48\linewidth}
	    \centering
        \includegraphics[width=0.9\linewidth]{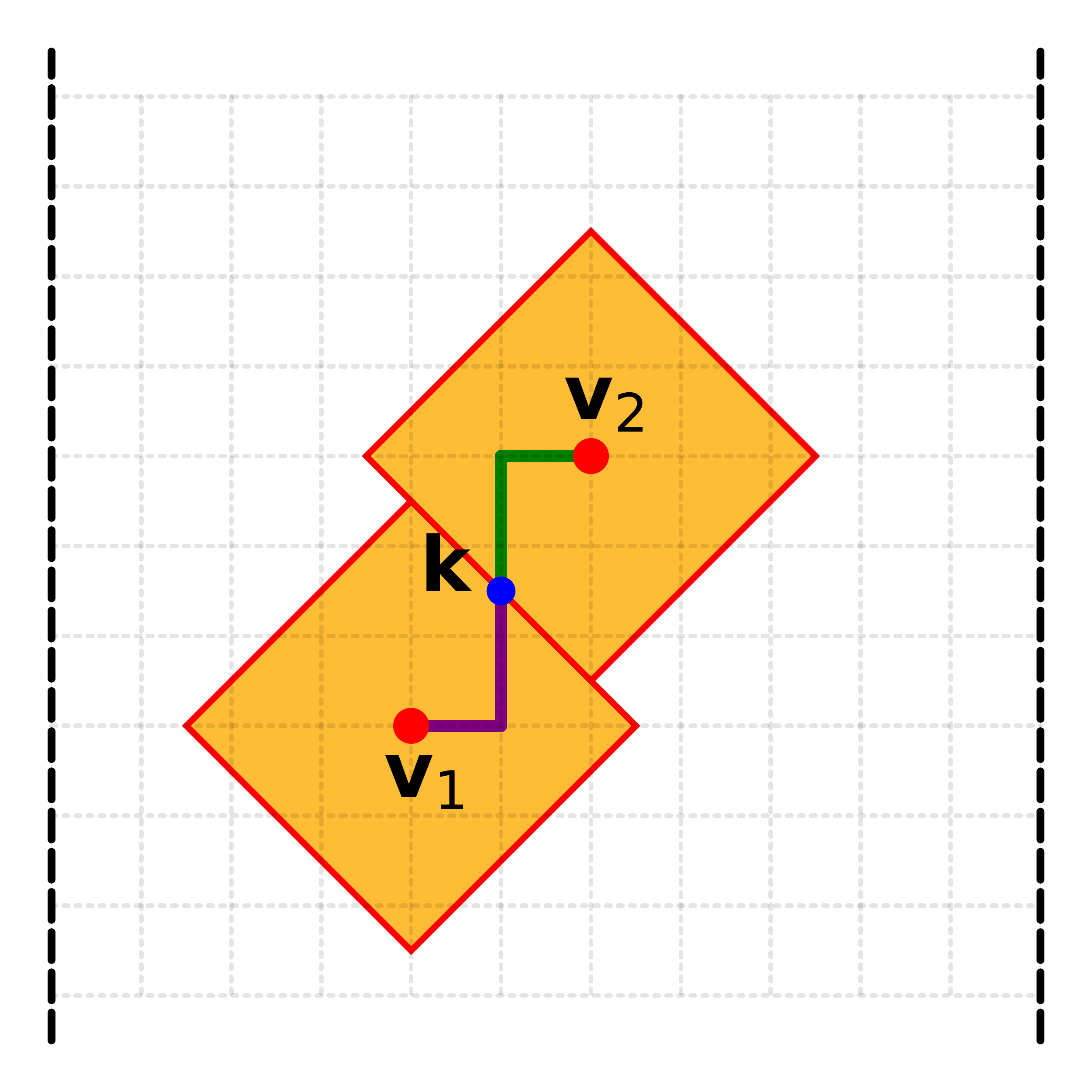}
        \caption{Decoding Graph}
        \label{fig:uf-equivalence-decoding-graph}
    \end{subfigure}
	\begin{subfigure}{.48\linewidth}
	    \centering
        \includegraphics[width=0.9\linewidth]{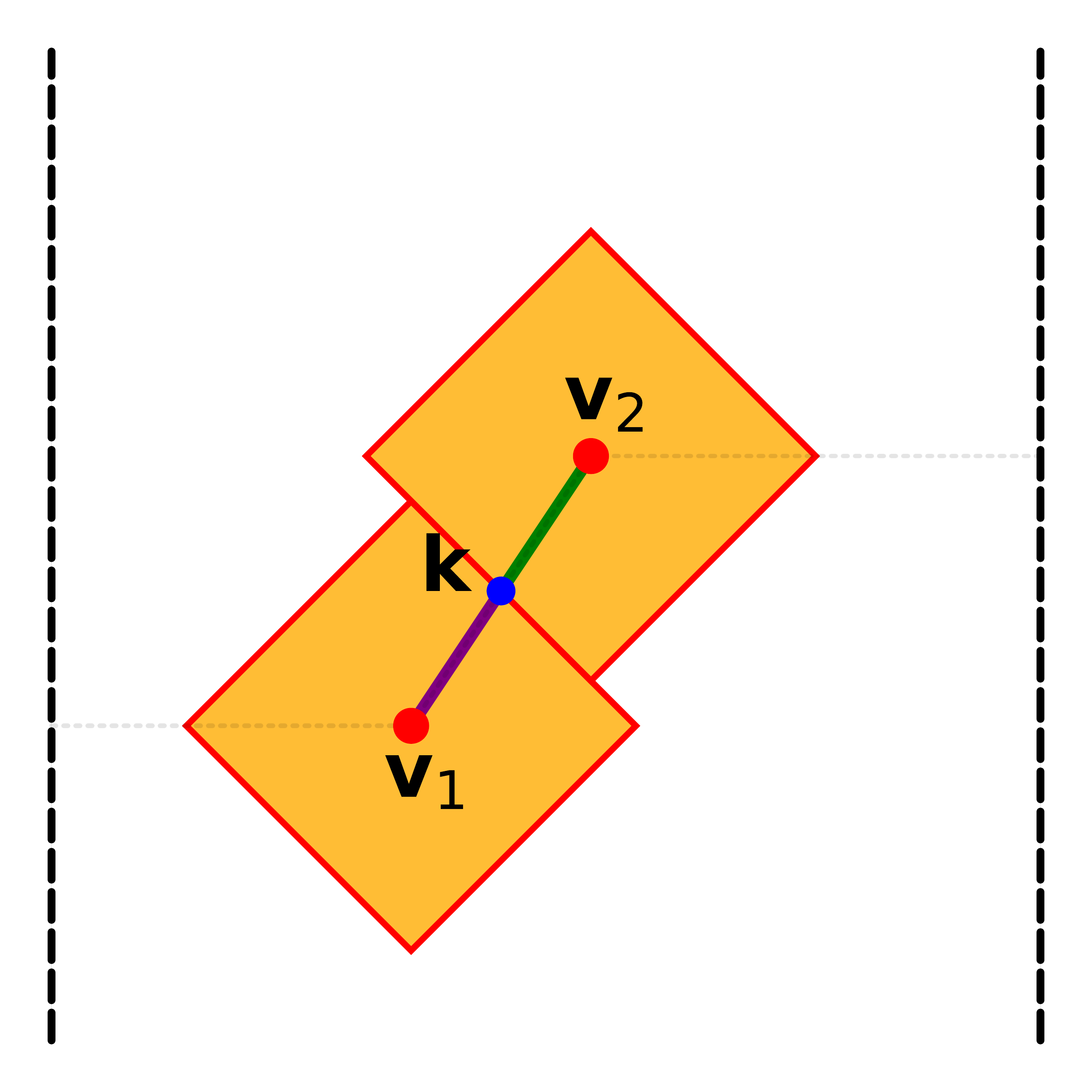}
        \caption{Syndrome Graph}
        \label{fig:uf-equivalence-syndrome-graph}
    \end{subfigure}
	\caption{Clusters on syndrome graph and decoding graph always touch at simultaneously.}
	\label{fig:uf-equivalence}
\end{figure}

While the original UF decoder works on the decoding graph~\cite{delfosse2021almost}, we show that the UF decoder works on the syndrome graph equivalently in terms of decoding accuracy.

The difference between the decoding graph and the syndrome graph is two-fold: the syndrome graph only have syndrome vertices $V^S$ while the decoding graph have all measurement vertices $V^D \supseteq V^S$; the syndrome graph is a complete graph where there is an edge between any pair of vertices $u, v \in V^S$.
Every edge $e = \langle u, v \rangle$ in the syndrome graph corresponds to the minimum-weight paths between $u$ and $v$ in the decoding graph.
We define the distance $d(u, v)$ between vertices $u, v$ in a graph as the weight of a minimum-weight path between them.
A point $k$ is either a vertex or a point on an edge. Similarly we can define $d(u, k)$ as the weight of a minimum-weight path from vertex $u$ to a point $k$.

In order to show that UF decoders on both graphs have the same decoding accuracy, we only need to show that the final clusters are the same, i.e. covering the same set of syndrome vertices.
The UF decoder logic is the same: it grows a cluster uniformly over all possible directions, and stops when it becomes even or touches a virtual boundary.
Using mathematical induction, if we can show that during the algorithm clusters always touch simultaneously on two graphs, then the final clusters are the same.
\autoref{fig:uf-equivalence} shows an example of clusters touching simultaneously on the decoding graph and the syndrome graph.

\section{Examples using the \diagram}

\begin{figure*}[ht]
    \renewcommand*\thesubfigure{(\arabic{subfigure})}  
	\centering
	\begin{subfigure}{.24\linewidth}
	    \centering
        \includegraphics[width=0.8\linewidth]{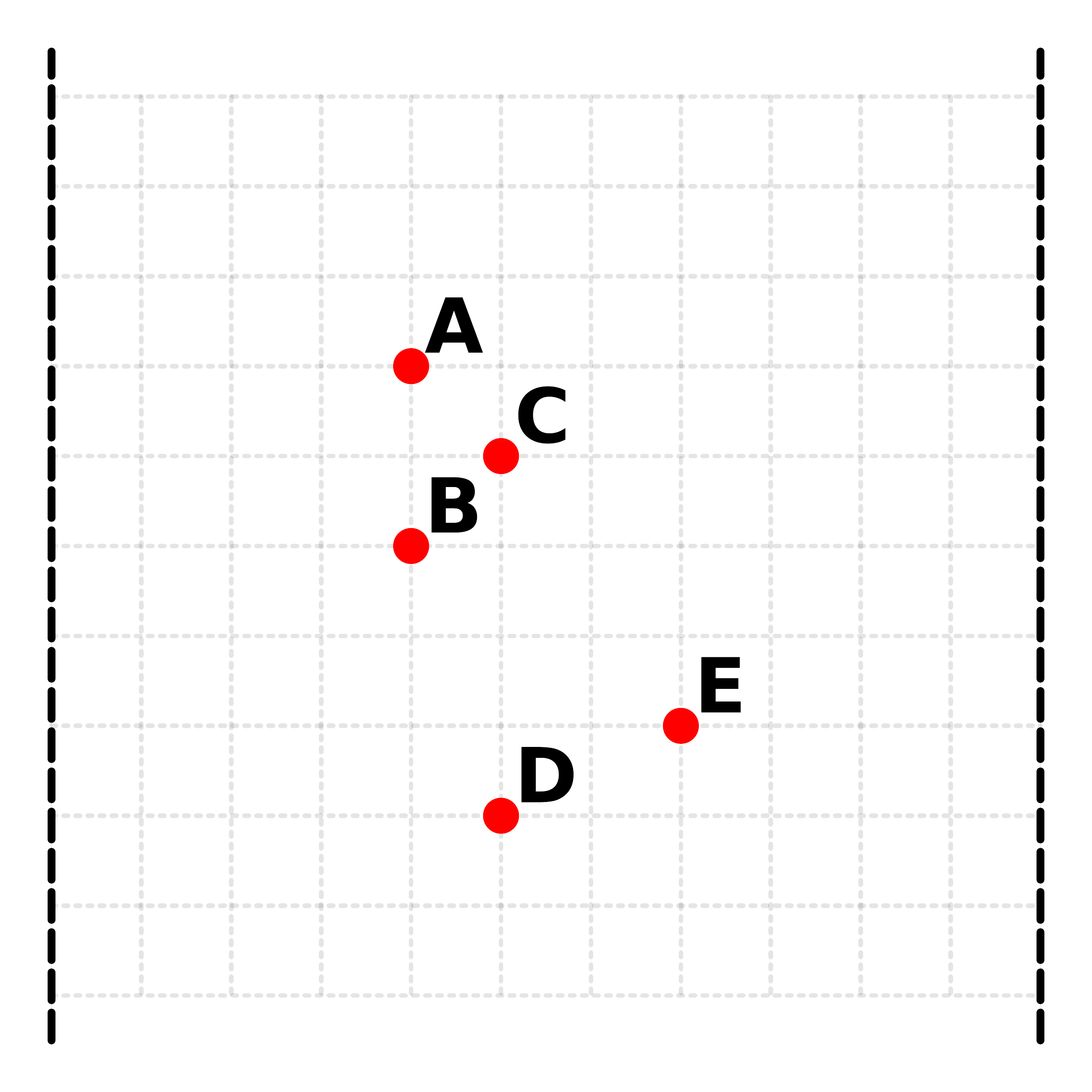}
        \caption{$\sum y = 0$}
        \label{fig:blossom-multi-grow-0}
    \end{subfigure}
	\begin{subfigure}{.24\linewidth}
	    \centering
        \includegraphics[width=0.8\linewidth]{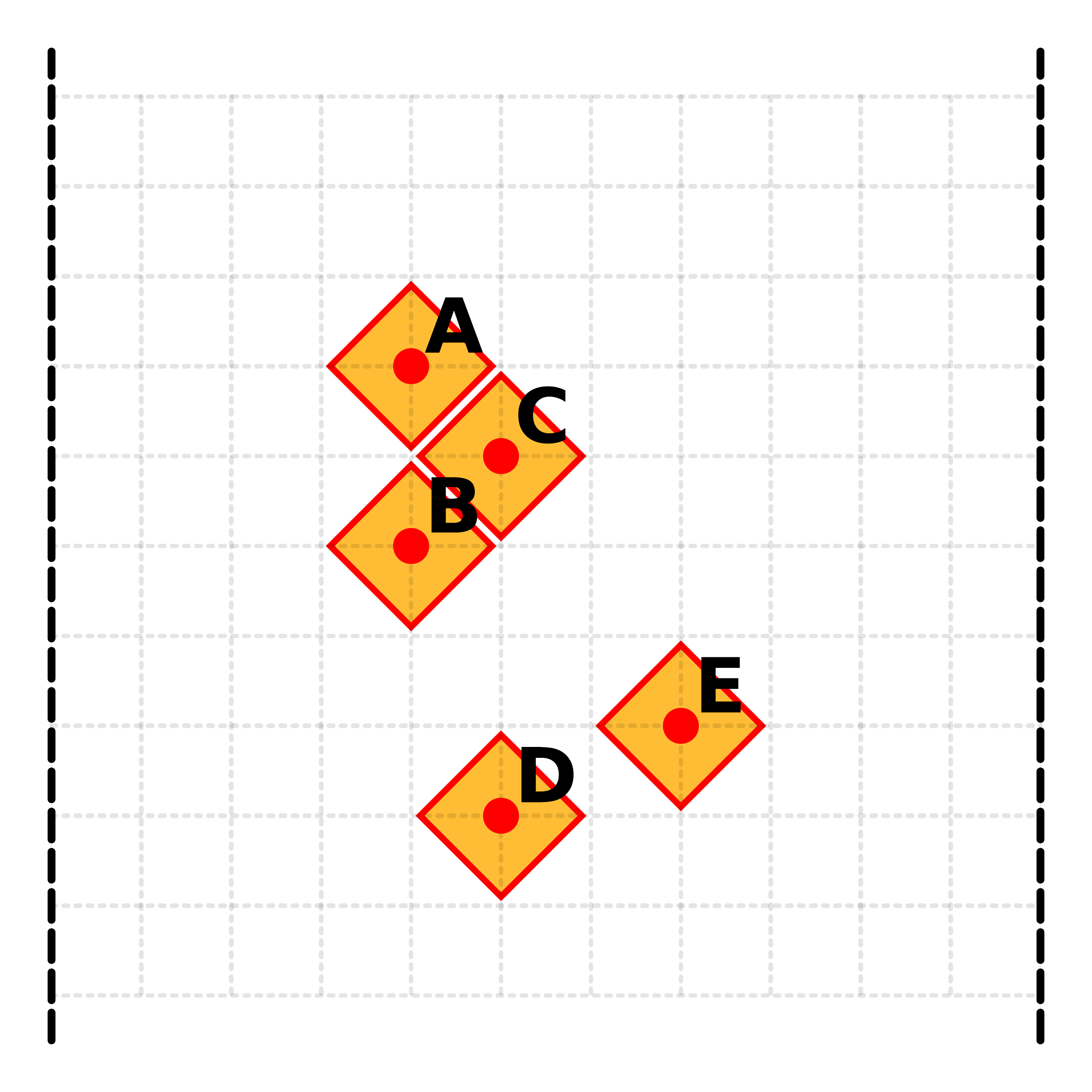}
        \caption{$\sum y = 4.5$}
        \label{fig:blossom-multi-grow-0.9}
    \end{subfigure}
	\begin{subfigure}{.24\linewidth}
	    \centering
        \includegraphics[width=0.8\linewidth]{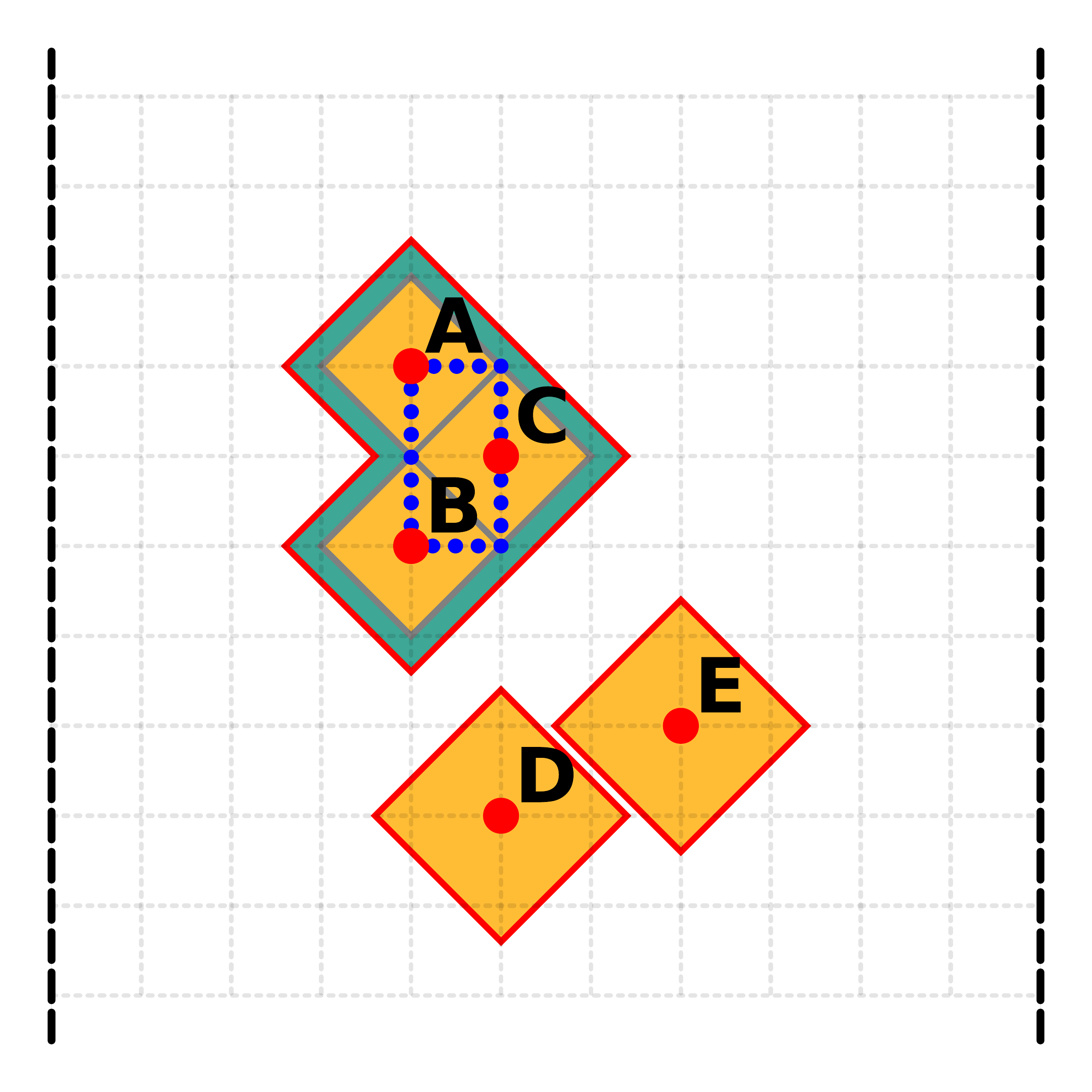}
        \caption{$\sum y = 6.2$}
        \label{fig:blossom-multi-grow-1.4}
    \end{subfigure}
	\begin{subfigure}{.24\linewidth}
	    \centering
        \includegraphics[width=0.8\linewidth]{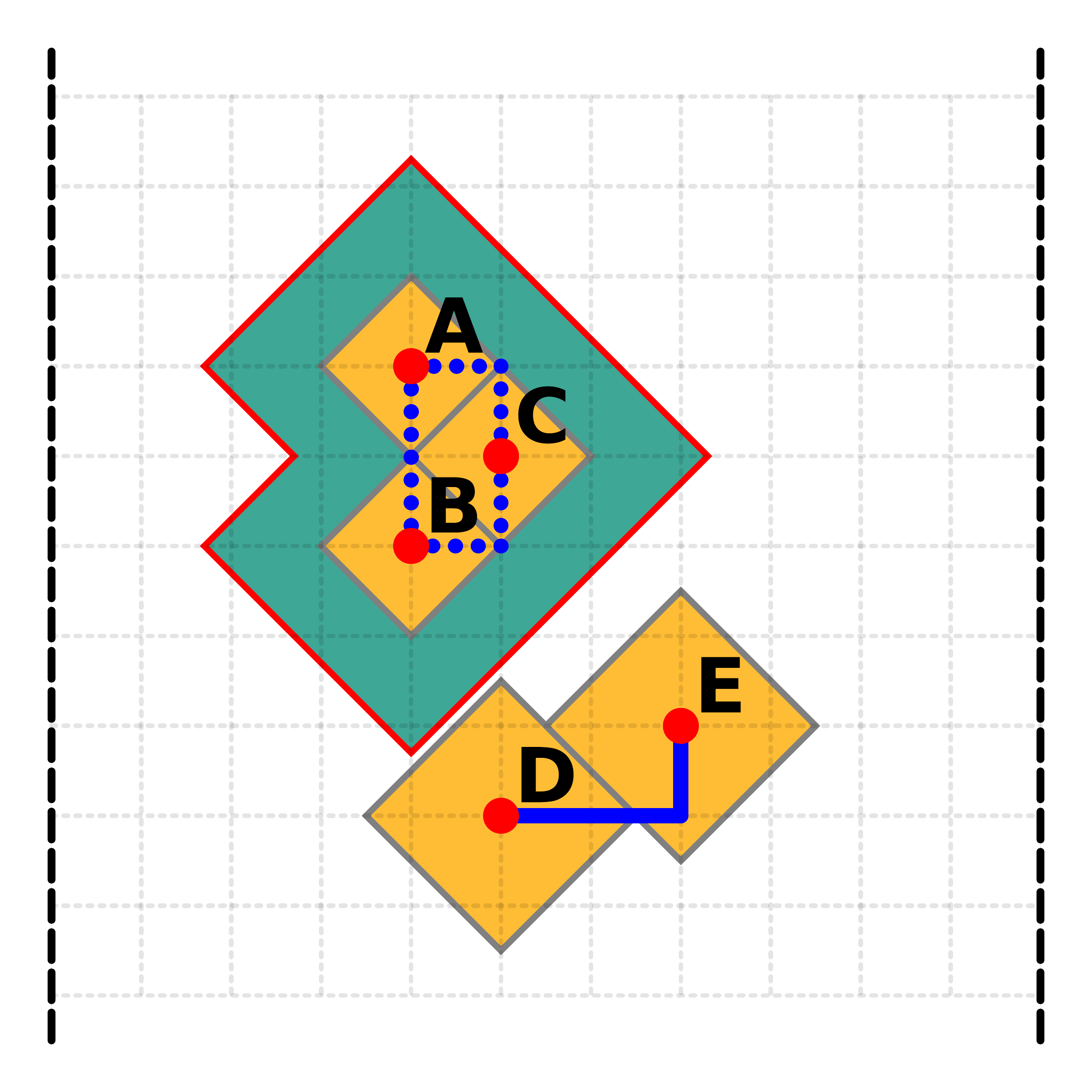}
        \caption{$\sum y = 7.3$}
        \label{fig:blossom-multi-grow-2.3}
    \end{subfigure}
	\begin{subfigure}{.24\linewidth}
	    \centering
        \includegraphics[width=0.8\linewidth]{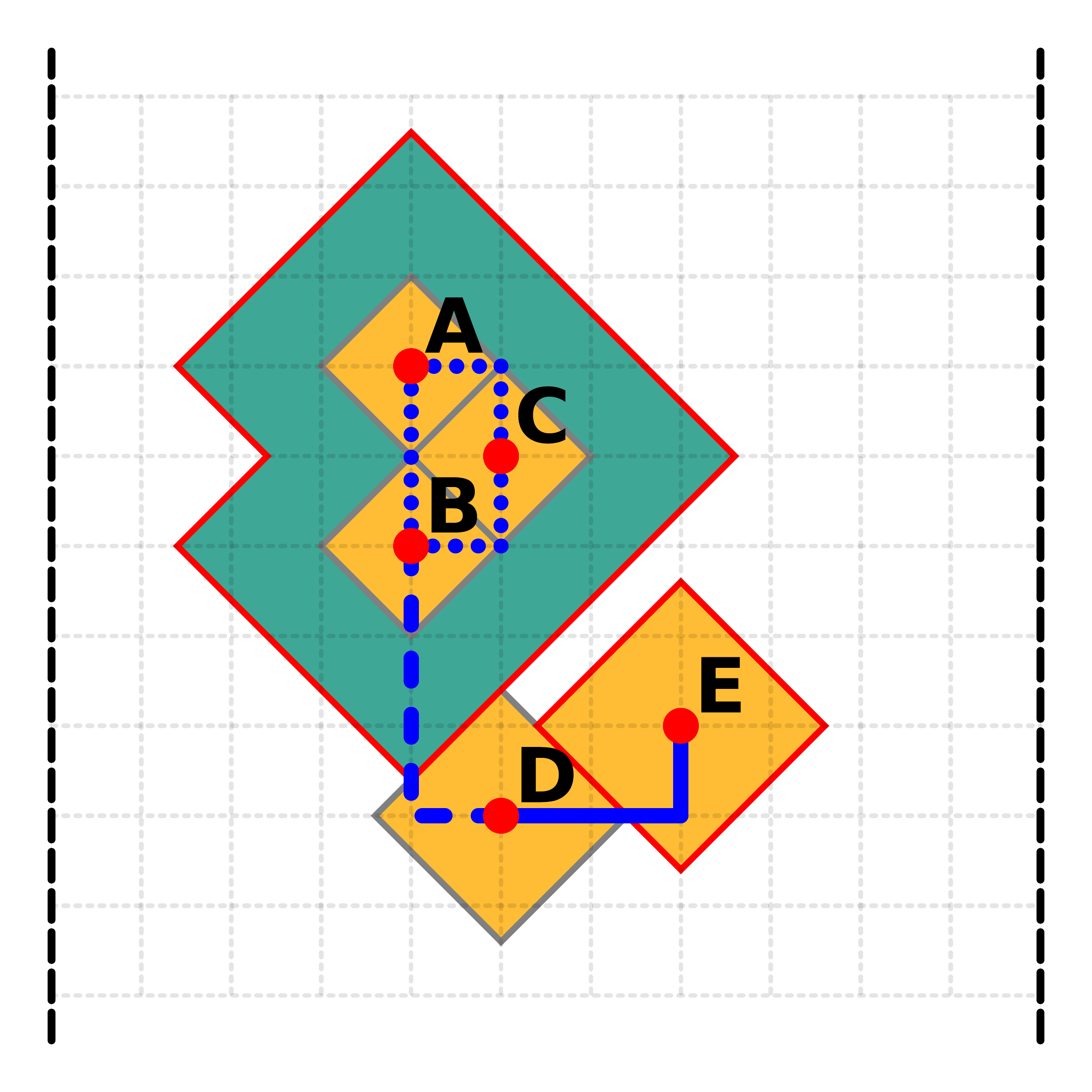}
        \caption{$\sum y = 7.6$}
        \label{fig:blossom-multi-grow-2.6}
    \end{subfigure}
	\begin{subfigure}{.24\linewidth}
	    \centering
        \includegraphics[width=0.8\linewidth]{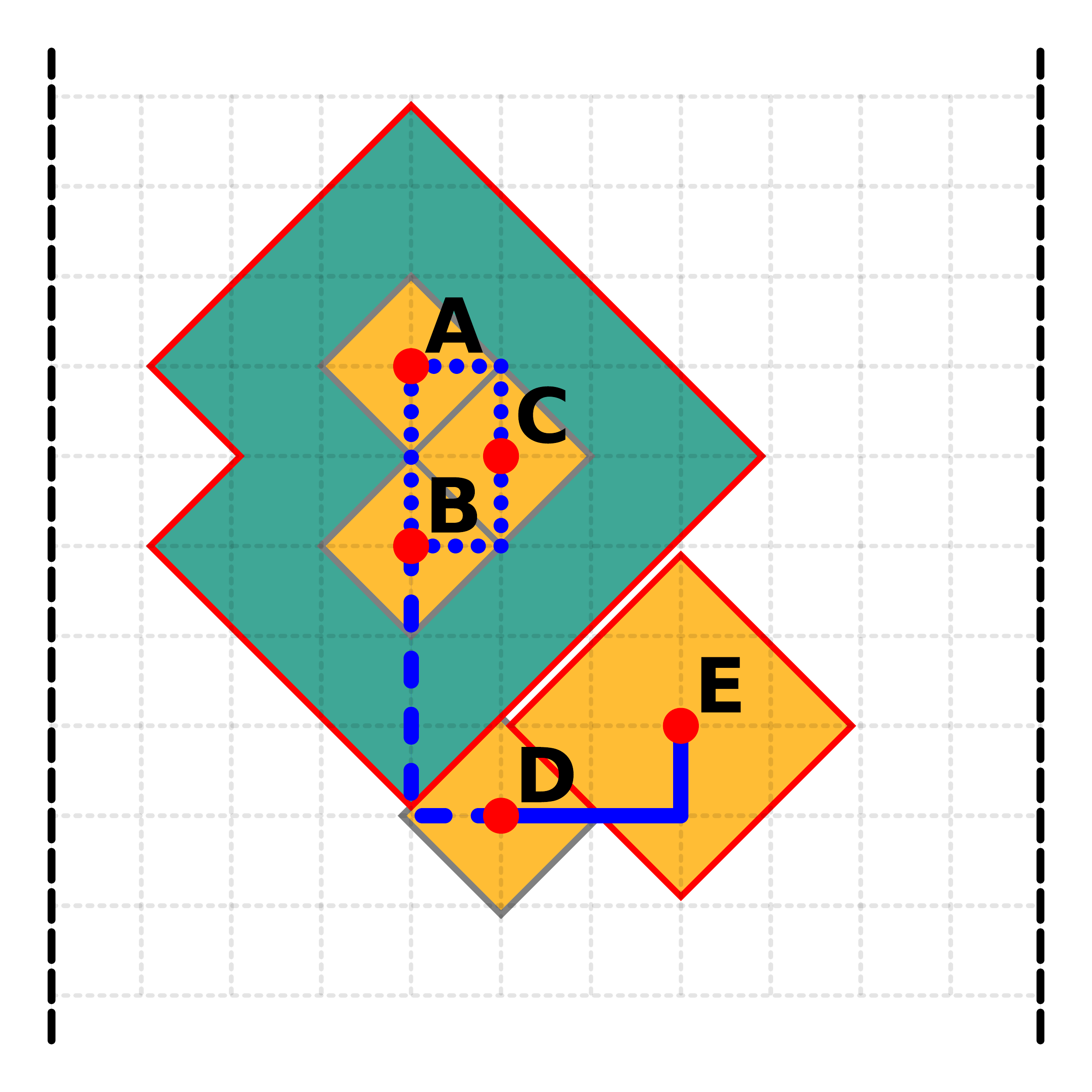}
        \caption{$\sum y = 7.9$}
        \label{fig:blossom-multi-grow-2.9}
    \end{subfigure}
	\begin{subfigure}{.24\linewidth}
	    \centering
        \includegraphics[width=0.8\linewidth]{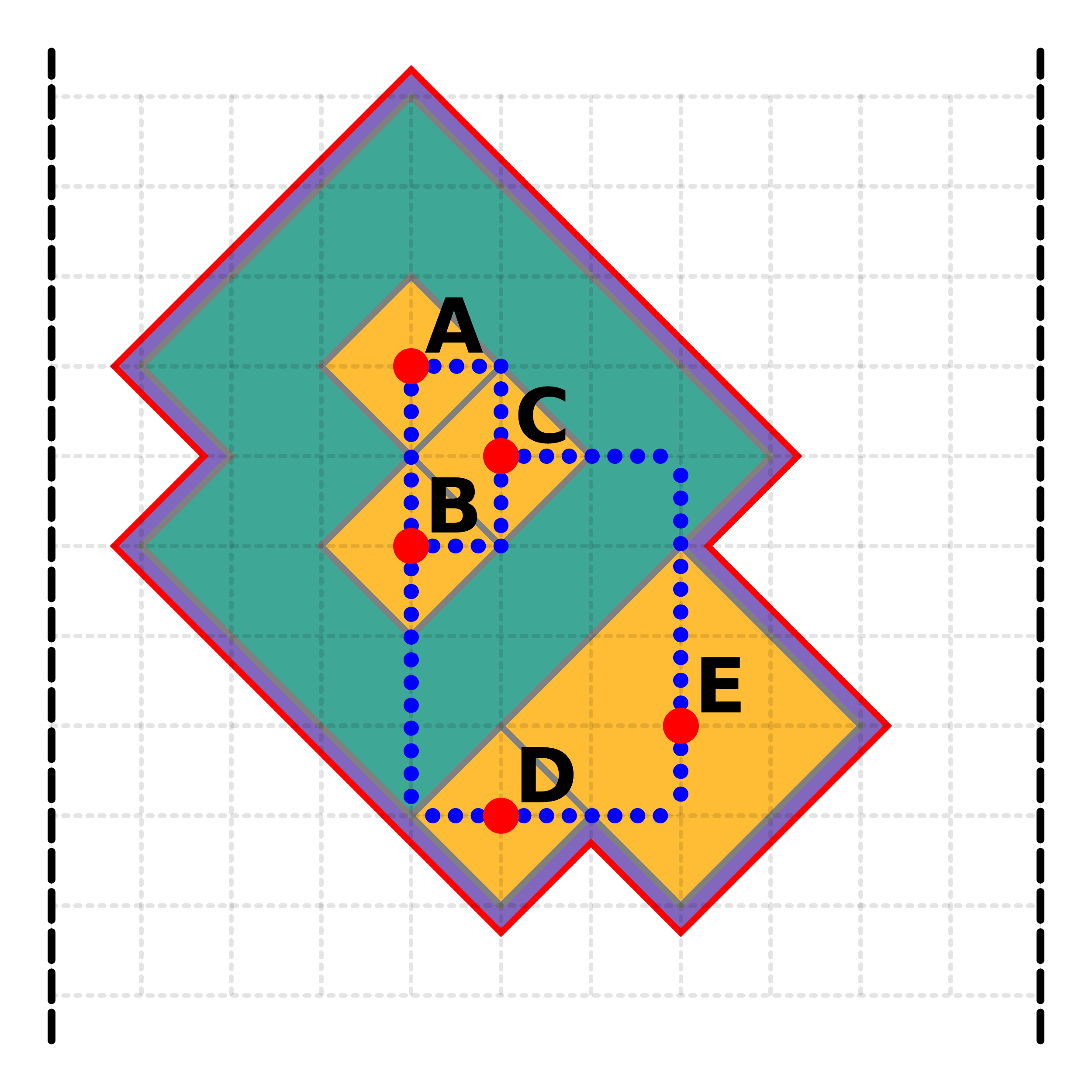}
        \caption{$\sum y = 8.3$}
        \label{fig:blossom-multi-grow-3.3}
    \end{subfigure}
	\begin{subfigure}{.24\linewidth}
	    \centering
        \includegraphics[width=0.8\linewidth]{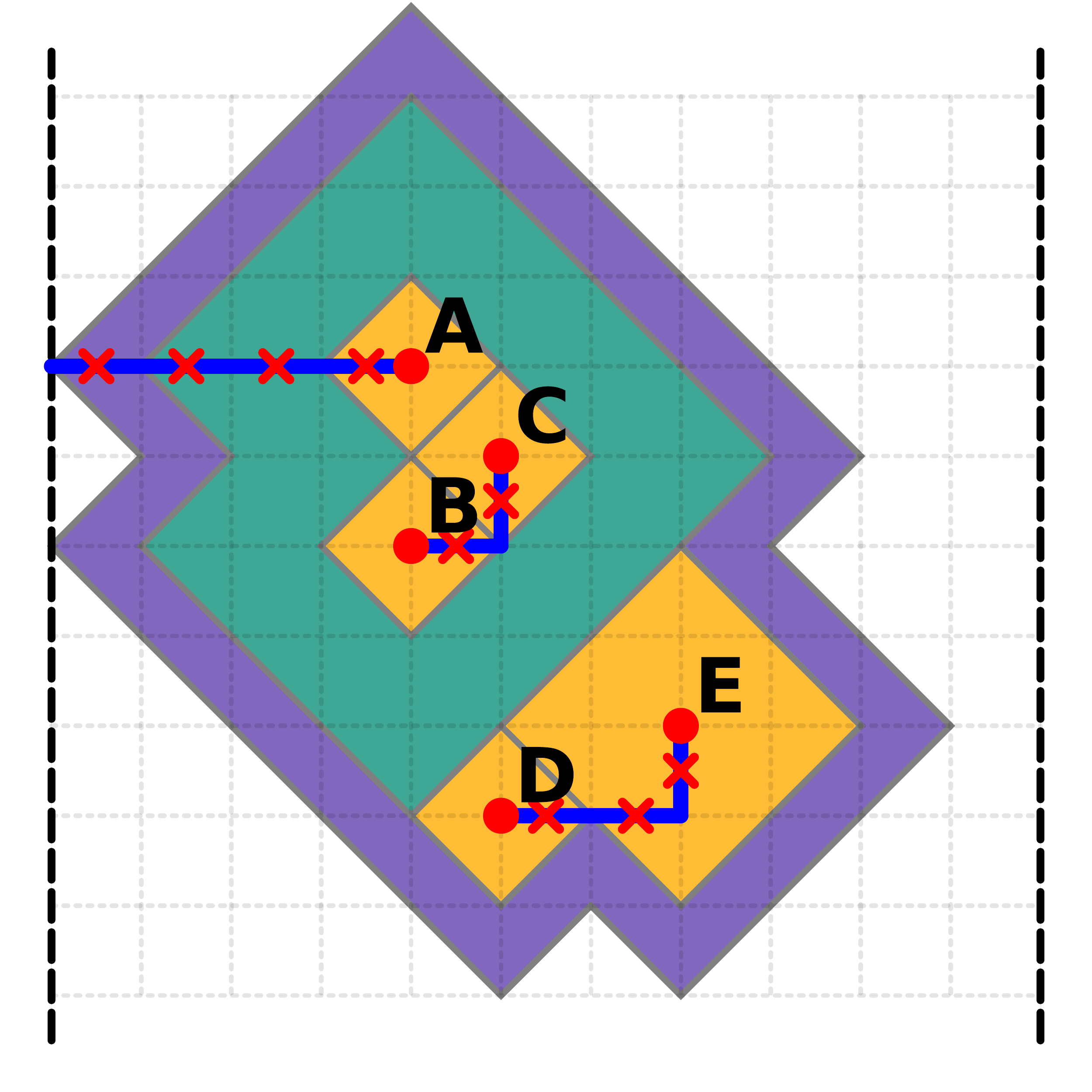}
        \caption{$\sum y = 9$}
        \label{fig:blossom-multi-grow-4}
    \end{subfigure}
	\caption{The blossom algorithm with multiple tree approach~\cite{kolmogorov2009blossom}. The primal variables are drawn as blue lines connecting vertices. The dual variables are drawn as the ``radius'' of the colored regions. When two regions touch, a tight edge is formed between two vertices from each region. (1) The initial state with all dual variables initialized to 0. (2) The dual variable of each vertex grows simultaneously. (3) When the regions of vertices $A$, $B$, $C$ touch each other, they form a \textit{blossom} marked in dotted blue lines. The dual variable of this blossom $y_{\{A,B,C\}}$ (green) grows but individual dual variables $y_A$, $y_B$ and $y_C$ stop growing. (4) When the regions of vertices $D$ and $E$ touch each other, they form a solved \cluster marked in solid blue line. The dual variables $y_D$, $y_E$ stop growing. (5) When the blossom $\{A,B,C\}$ touches the solved cluster $\{D,E\}$, it forms an alternating tree marked in dashed and solid blue lines. (6) In this alternating tree, $y_{\{A,B,C\}}$ and $y_E$ grows as usual, but $y_D$ shrinks at the same speed. In this way, they still keep touch with each other. (7) A bigger blossom is formed and its dual variable $y_{\{A,B,C,D,E\}}$ (purple) starts growing. Individual dual variables $y_{\{A,B,C\}}$, $y_D$ and $y_E$ stop growing. (8) This blossom touches a left virtual boundary vertex and breaks into three temporary matches: $A$ matches to the left virtual boundary vertex, $B$ matches to $C$, $D$ matches to $E$. The cluster is now solved, so the blossom algorithm terminates. The minimum-weight perfect matching is the collection of all those matchings with only tight edges.}
	\label{fig:blossom_multi_grow}
\end{figure*}

\begin{figure*}[ht]
    \renewcommand*\thesubfigure{(\arabic{subfigure})}  
    	\centering
	\begin{subfigure}{.24\linewidth}
	    \centering
        \includegraphics[width=0.8\linewidth]{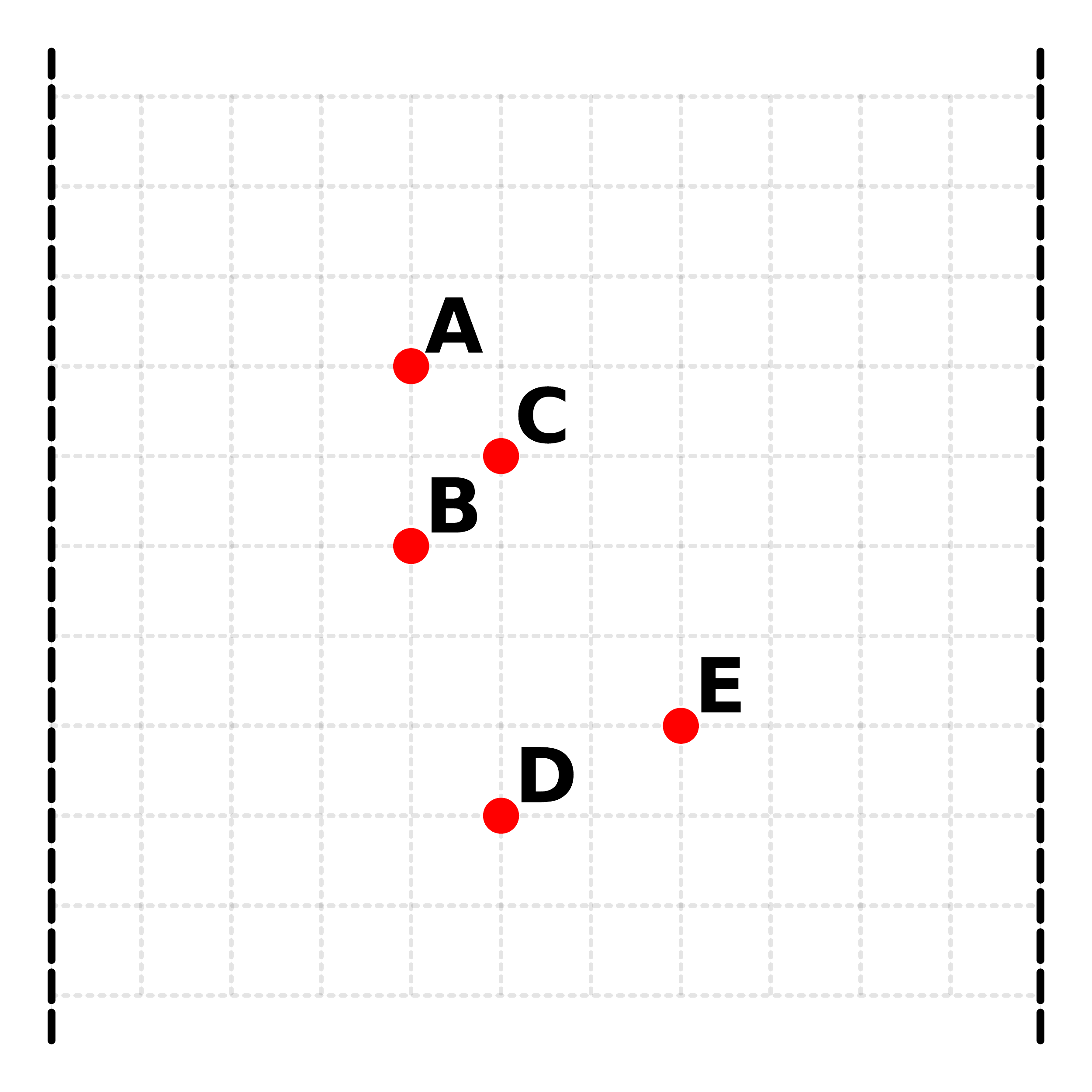}
        \caption{}
        \label{fig:union_find_multi_grow_0}
    \end{subfigure}
	\begin{subfigure}{.24\linewidth}
	    \centering
        \includegraphics[width=0.8\linewidth]{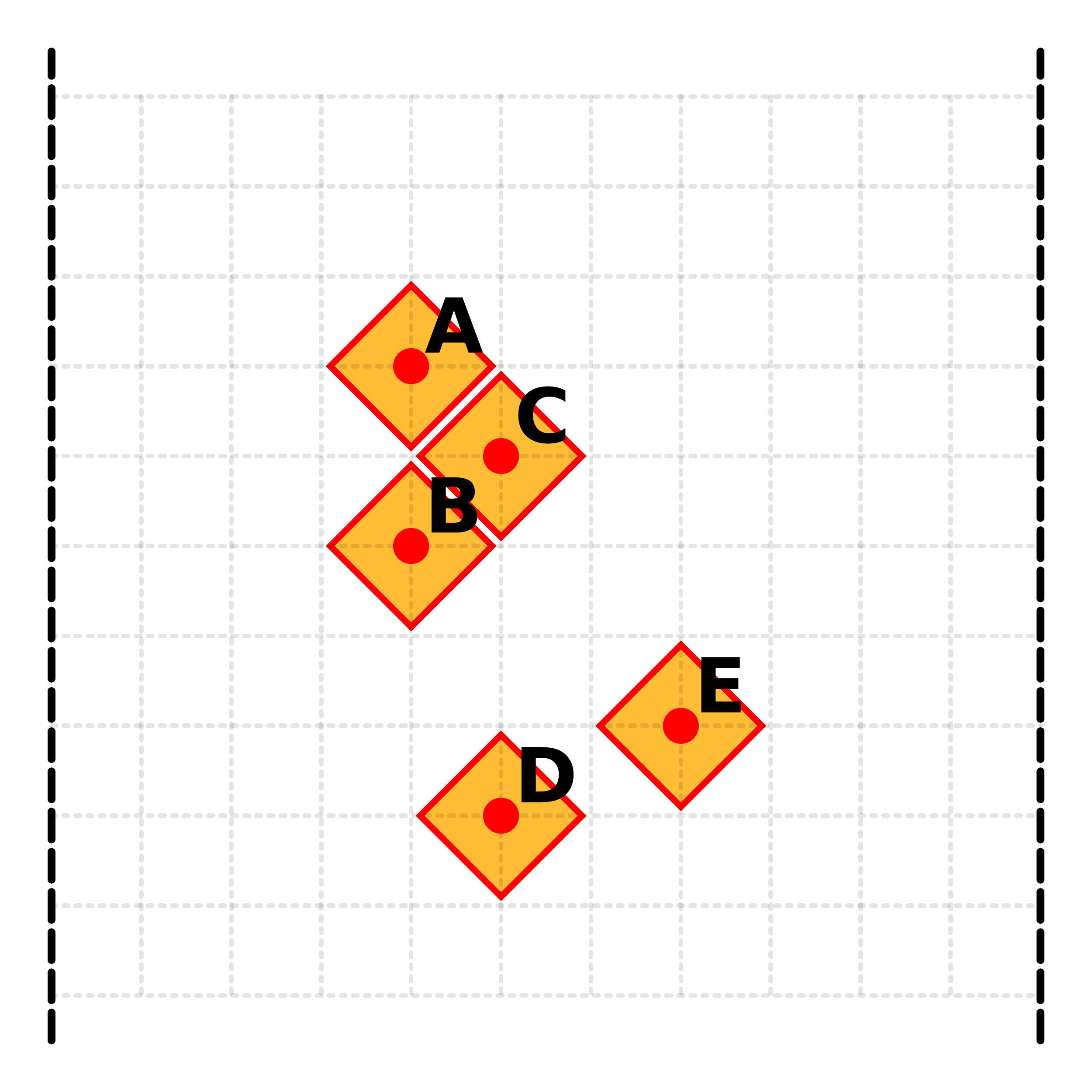}
        \caption{}
        \label{fig:union_find_multi_grow_0.9}
    \end{subfigure}
	\begin{subfigure}{.24\linewidth}
	    \centering
        \includegraphics[width=0.8\linewidth]{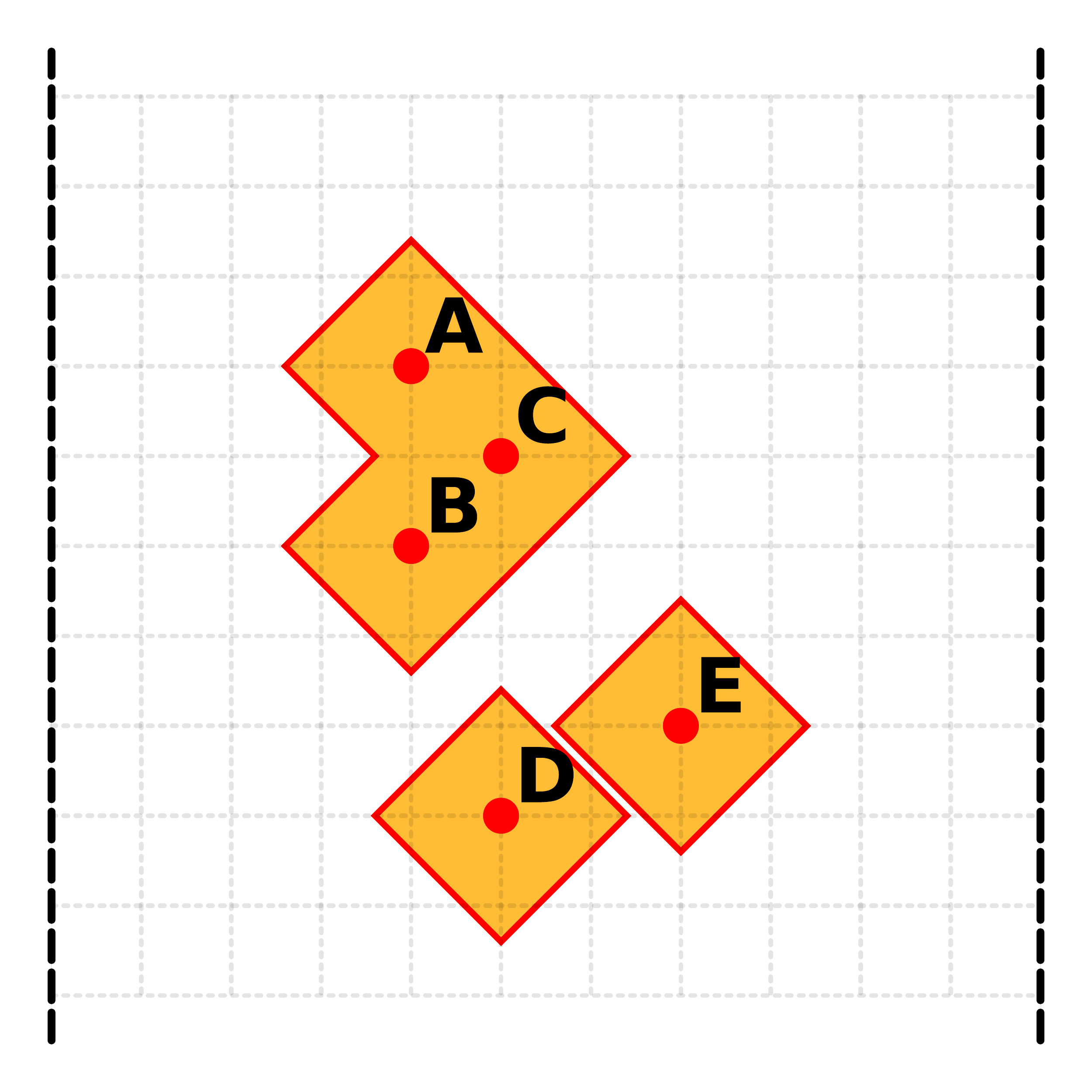}
        \caption{}
        \label{fig:union_find_multi_grow_1.4}
    \end{subfigure}
	\begin{subfigure}{.24\linewidth}
	    \centering
        \includegraphics[width=0.8\linewidth]{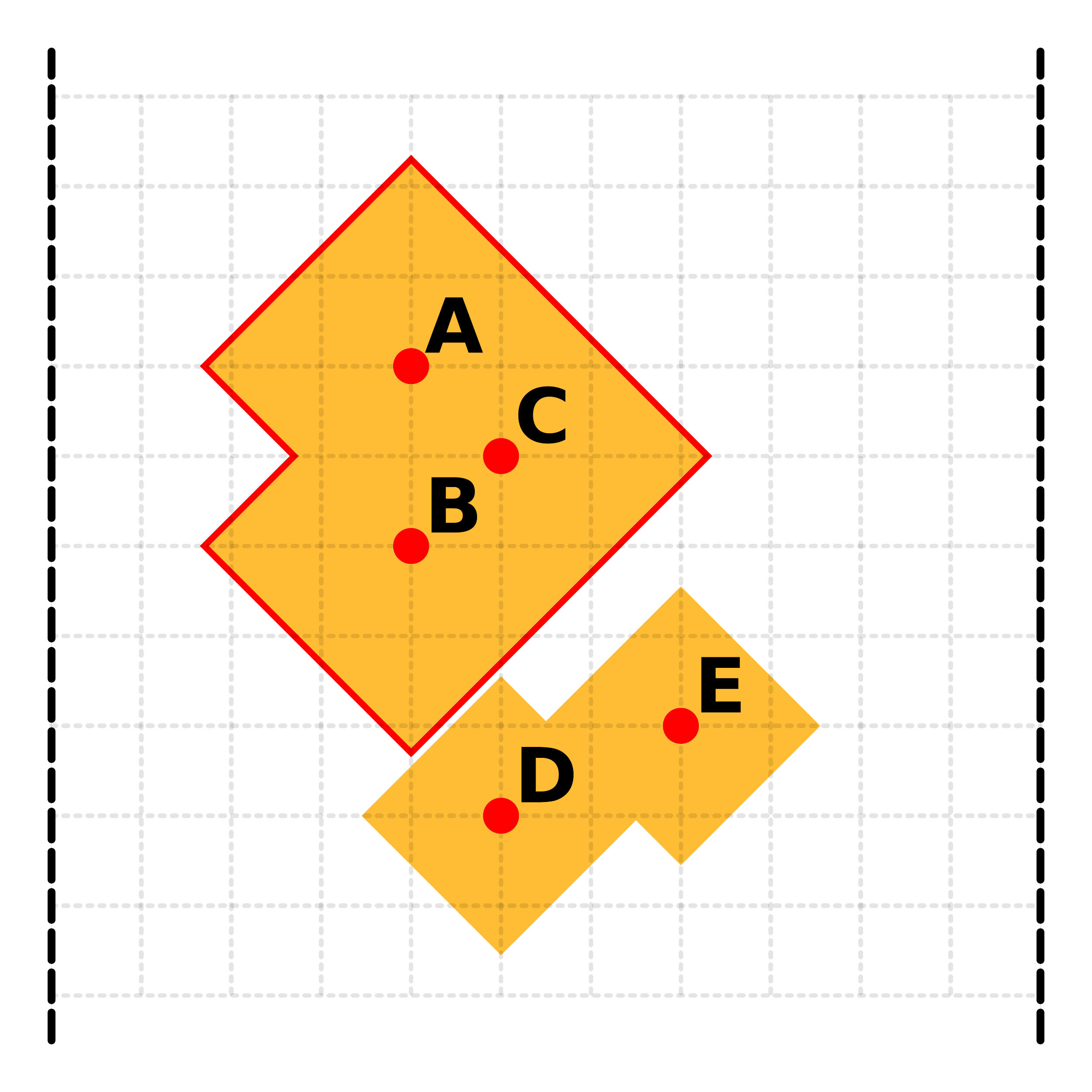}
        \caption{}
        \label{fig:union_find_multi_grow_2.3}
    \end{subfigure}
	\begin{subfigure}{.24\linewidth}
	    \centering
        \includegraphics[width=0.8\linewidth]{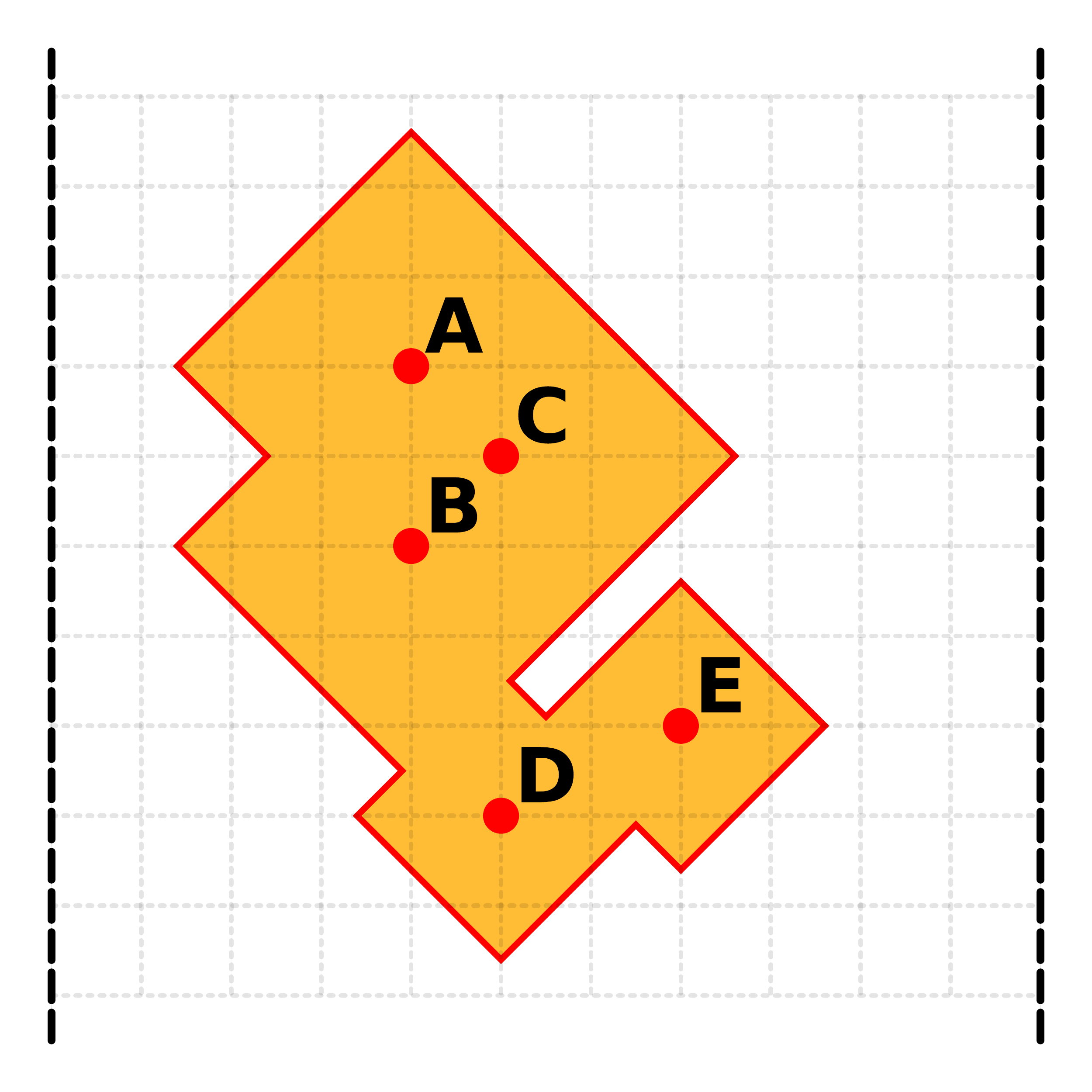}
        \caption{}
        \label{fig:union_find_multi_grow_2.6}
    \end{subfigure}
	\begin{subfigure}{.24\linewidth}
	    \centering
        \includegraphics[width=0.8\linewidth]{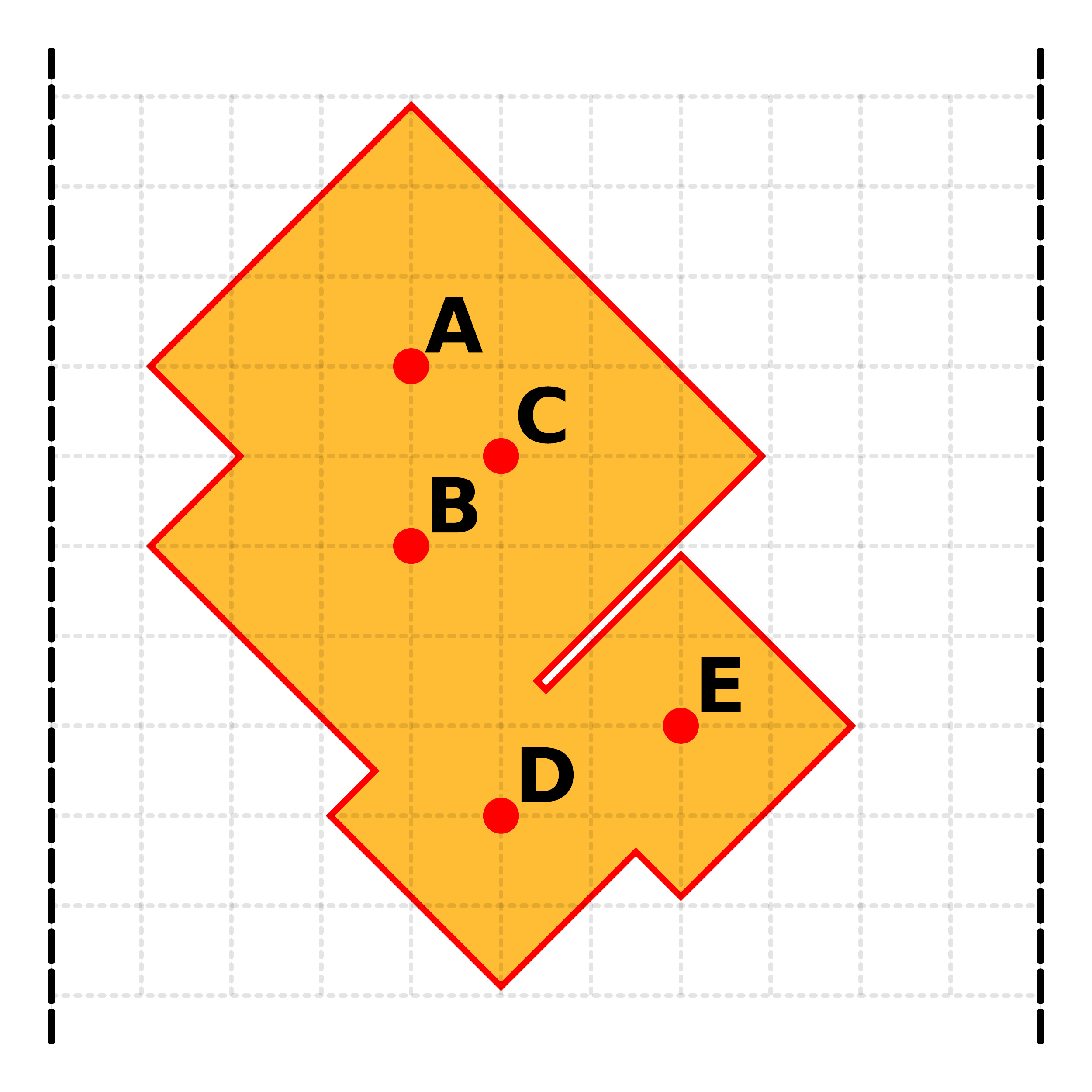}
        \caption{}
        \label{fig:union_find_multi_grow_2.9}
    \end{subfigure}
	\begin{subfigure}{.24\linewidth}
	    \centering
        \includegraphics[width=0.8\linewidth]{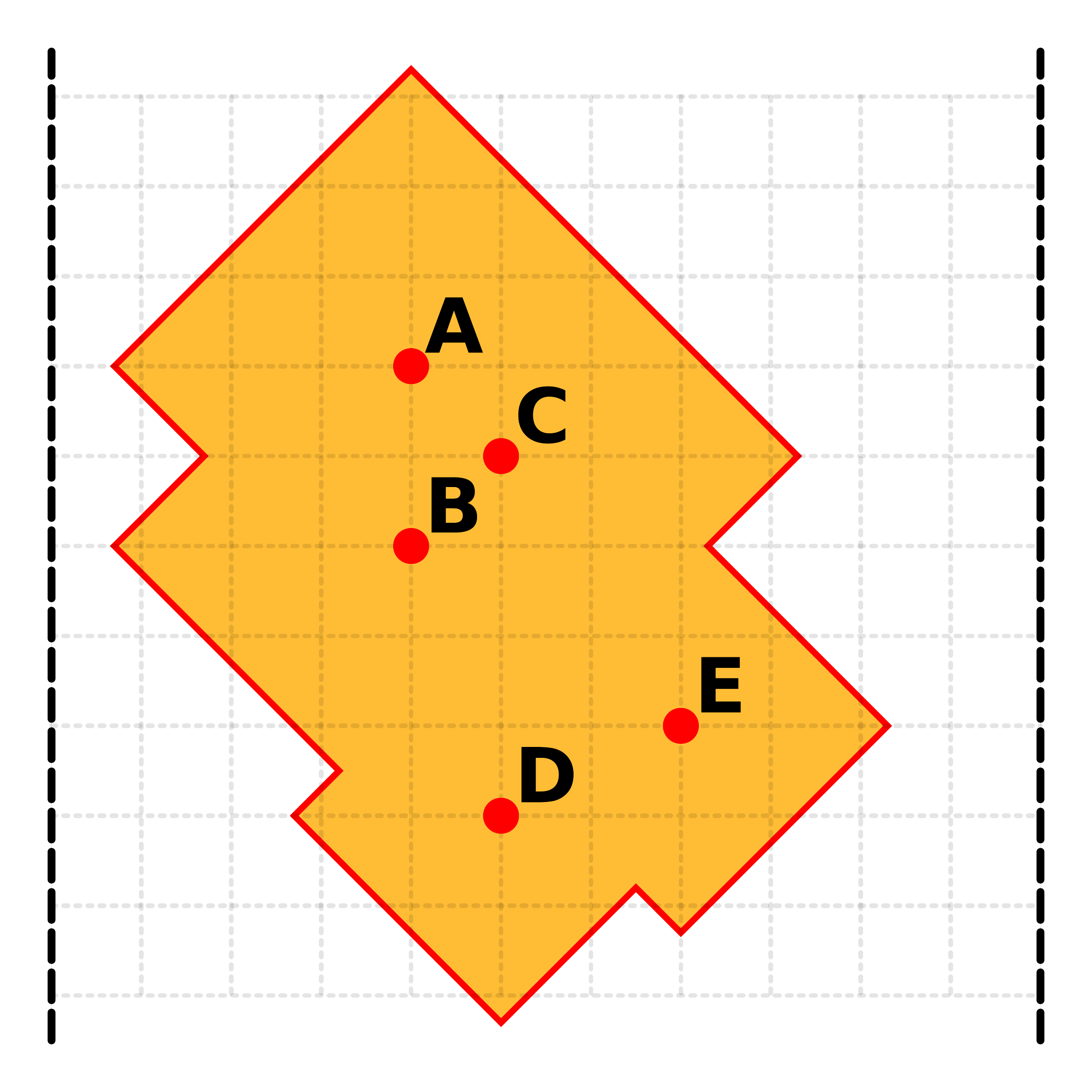}
        \caption{}
        \label{fig:union_find_multi_grow_3.3}
    \end{subfigure}
	\begin{subfigure}{.24\linewidth}
	    \centering
        \includegraphics[width=0.8\linewidth]{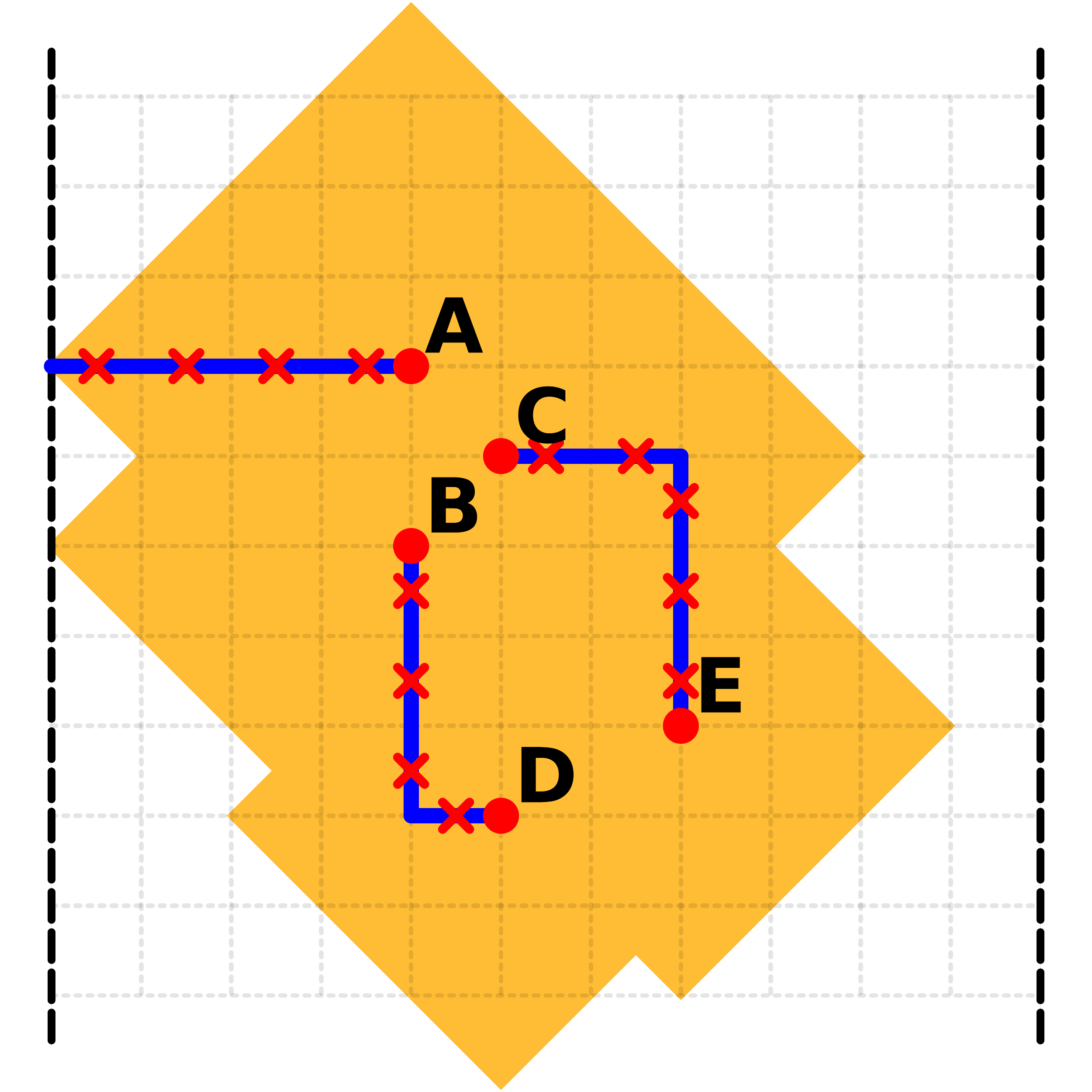}
        \caption{}
        \label{fig:union_find_multi_grow_4}
    \end{subfigure}
	\caption{The union-find decoder. A \textit{grown edge} is fully covered by regions (yellow). A \textit{half-grown edge} is partly covered by regions. An \textit{unoccupied edge} is not covered by any region. (1) Initially all edges are unoccupied. (2) Each vertex is an \textit{odd cluster} and grows uniformly. Odd (even) cluster consists of odd (even) number of vertices. (3) Clusters merge together. The new cluster $\{A,B,C\}$ is still an odd cluster and keeps growing. (4) Two odd clusters $D$ and $E$ merge into an \textit{even cluster} and stop growing. (5) The odd cluster $\{A,B,C\}$ merges with the even cluster $\{D,E\}$ and becomes a bigger odd cluster $\{A,B,C,D,E\}$. This bigger cluster grows uniformly even though the cluster $\{D,E\}$ has been stopped for a while. (6)(7)(8) The odd cluster $\{A,B,C,D,E\}$ keeps growing until it touches a left virtual boundary vertex and terminates. After all clusters stop growing, the union-find decoder applies the \textit{peeling algorithm} to find an error pattern that generates this syndrome using single-qubit errors only inside each cluster. This error pattern corresponds to a perfect matching marked in blue lines in (8), though generally not a minimum-weight perfect matching. Note that each sub-figure corresponds to one in \autoref{fig:blossom_multi_grow}, with similar shape of  \regions.}
	\label{fig:union_find_multi_grow}
\end{figure*}

\end{document}